\documentclass[aps,prx,amsmath,amsfonts,twocolumn]{revtex4-1}%
\usepackage{mathtools}
\usepackage{verbatim}
\usepackage{hyperref}
\usepackage{epsfig,color,amsfonts,amsmath,amsthm,comment,natbib}
\usepackage{color}
\usepackage{graphicx}
\usepackage{braket}
\usepackage{amsthm}
\usepackage{amssymb}
\usepackage{amsmath}
\usepackage{ulem}
\usepackage{tikz}
\usepackage{subcaption}

\DeclareMathOperator{\Tr}{Tr}

\providecommand{\del}[0]{\ensuremath{\vec{\nabla}}}

\newcommand{\norm}[1]{\left|\left|#1\right|\right|}

\begin{document}

\title{Stirring by staring: measurement induced chirality}

\author{Matthew Wampler$^{1}$}
\email{mbw5kk@virginia.edu}
\author{Brian J. J. Khor$^{1}$, Gil Refael$^{2,3}$}
\author{Israel Klich$^{1}$}
\email{ik3j@virginia.edu}
\affiliation{$^1$Department of Physics, University of Virginia, Charlottesville, Virginia 22903, USA
}
\affiliation{$^2$Department of Physics, California Institute of Technology, Pasadena CA 91125, USA}
\affiliation{$^3$Institute for Quantum Information and Matter,
California Institute of Technology, Pasadena CA 91125, USA}
\begin{abstract}
In quantum mechanics, the observer necessarily plays an active role in the dynamics of the system, making it difficult to probe a system without disturbing it. Here, we leverage this apparent difficulty as a tool for driving an initially trivial system into a chiral  phase. In particular, we show that by utilizing a pattern of repeated occupation measurements we can produce chiral edge transport of fermions hopping on a Lieb lattice.  The procedure is similar in spirit to the use of periodic driving to induce chiral edge transport in Floquet topological insulators, while also exhibiting novel phenomena due to the non-unitary nature of the quantum measurements.  We study in detail the dependence of the procedure on measurement frequency, showing that in the Zeno limit the system can be described by a classical stochastic dynamics, yielding protected transport. As the frequency of measurements is reduced, the charge flow is reduced and vanishes when no measurements are done. 
\end{abstract}


\maketitle

\section{Introduction}
One of the most exciting goals of the field of quantum dynamics is to be able to control the microscopic motion of particles in a reliable and universal way. Floquet engineering, coupled with our knowledge of topological quantum phases, presented one such route and brought about new paradigms for the quantum control of atomic and electronic motion. 
A periodic modulation of the Hamiltonian was shown to induce Chern bands in non-topological semiconductors as well as graphene, and this remarkable feat was observed in a variety of solid state and atomic systems \cite{lindner2011floquet,McIver_2019,Nuske_2020}.

The range of drive-induced topological phases kept growing over the past decade to include states with no static analogs. A prominent example is the anomalous Floquet Anderson insulator \cite{Titum_2015,2020PhRvB.101d1403K,2019PhRvB..99s5133N,Po_2016}. In this 2D phase, a chiral edge state emerges alongside completely trivial bulk bands in stark contrast to standard topological edge states which are spectrally connected to bulk bands. Thus, such an insulator avoids issues associated with Fermion anomalies. The trick behind this phase is a Floquet Hamiltonian modulation which alters the hopping along a square lattice in a sequence that stirs the particles \cite{PhysRevX.3.031005} in such a way that bulk motion is cancelled and edge states emerge. 

An additional tool for control, however, is measurement (e.g. \onlinecite{2020PhRvR...2c3347R,2020PhRvL.124d0401L}). 'Dark state' engineering was explored as a means to stabilize a variety of phases through measurement or decay processes that eliminate unwanted elements in the wave function in order to stabilize a desired steady state \cite{2021arXiv210508076M,2021PhRvR3b3200S,Budich_2015,Bardyn_2013,Tomadin_2012}. The challenge in this approach is to engineer the necessary projectors. The combination of periodic driving and dissipation has also been discussed \cite{2021arXiv210705669S,2018PhRvB..98t5417Z}. 
More recently, it was discovered that a combination of unitary evolution and measurement could actually induce a transition between highly entangled quantum states into low entanglement classical-looking states at high measurement frequency \cite{2021PhRvB.103j4306L,2021arXiv210208381B,Bao_2020,2019PhRvB..99v4307C,2020PhRvX..10d1020G,2020PhRvB.101j4302J,2019PhRvB.100f4204S,2021arXiv210703393Z}.  
The study of the competing effects of projective measurement and unitary evolution has also been intensely researched in the context of quantum circuit models \cite{skinner2019measurement,Yaodong2018, li2019measurement, Gullans2020Probe, Choi2020,Vasseur2019,Jian2020,sang2020measurement,Rossini2020,Zabalo2020,Fan2021, ippoliti2021entanglement, lavasani2021measurement, sang2020entanglement, shi2020entanglement}. 
The physics of measurement-induced phase transitions has been studied in the context of measurement protected quantum orders \cite{sang2020measurement}, symmetry-protected topological (SPT) phases \cite{lavasani2021measurement}, geometric phase \cite{gebhart2020topological}, many-body localization \cite{lunt2020measurement}, and various aspects of entanglement measures \cite{skinner2019measurement, li2019measurement, chan2019unitary, sang2020entanglement, ippoliti2021entanglement, shi2020entanglement}. There are also recent works which study the entanglement transitions with measurement and unitary evolution for free fermions hopping on a 1D chain \cite{Cao2019,alberton2020trajectory, 2021arXiv210208381B}. 
In \cite{klich2019closed} the competing effects of unitary evolution and  measurements were studied using a closed hierarchy approach.  
This method was used to describe non-equilibrium steady states of current  \cite{klich2019closed} as well as density fluctuations (quantum wakes) following a moving particle detector and other disturbances  \cite{wampler2021quantum}.

In this manuscript we combine these developments to show that measurements can stabilize protected 
edge transport. 
All that is needed is a sequence of local occupation measurements which serve to herd particles into circular orbits. These circular orbits then play a somewhat similar role to the semi-classical orbits used to illustrate the quantum hall effect \cite{Klitzing1980Hall,Jain2007Hall} where particles take closed trajectories in the bulk while the presence of an edge induces chiral motion via 'skipping orbits' \footnote{ The analogy is incomplete, however, in the Zeno limit of rapid measurements where the measurement induced chirality is a completely classical effect (as opposed to the quantum hall effect where coherence plays a key role).  The effect, however, does persist in the regime of less frequent measurements where the consequences of quantum coherence become important}.  The result, so called stirring by staring, combines the pioneering ideas of dark-state engineering with Floquet engineering to generate exotic protected edge dynamics. 
As a simple example, we show how this can be accomplished on a Lieb lattice where chirality is achieved via an 8-step measurement pattern.


We show that our measurement scheme, illustrated in Fig. \ref{fig:meas_protocol} and explained in detail below, yields no net transport of particles in the bulk of the lattice. However, when the system has an edge, it will induce movement of particles along the edge, Fig. \ref{fig:full loop}. We explore the evolution of particle density in the system using the closed hierarchy method \cite{klich2019closed} both by direct numerical simulation as well as by analytically studying the Zeno limit of rapid measurements, where the transport can be conveniently described as a stochastic process, and corrections to this process in the near-Zeno limit. In this regime, we prove that the boundary transport is protected from a wide range of edge perturbations  including random potentials, hopping energies, edge deformations, and site removal. It is critical to note that such protection cannot be achieved in a 1D system (with a strictly local Hamiltonian), where a removal of a small set of sites can simply disconnect the system into disjoint parts with no possibility of transport.  

\section{The protocol}

\begin{figure}[h]
    \centering
    \includegraphics[width=0.5\textwidth]{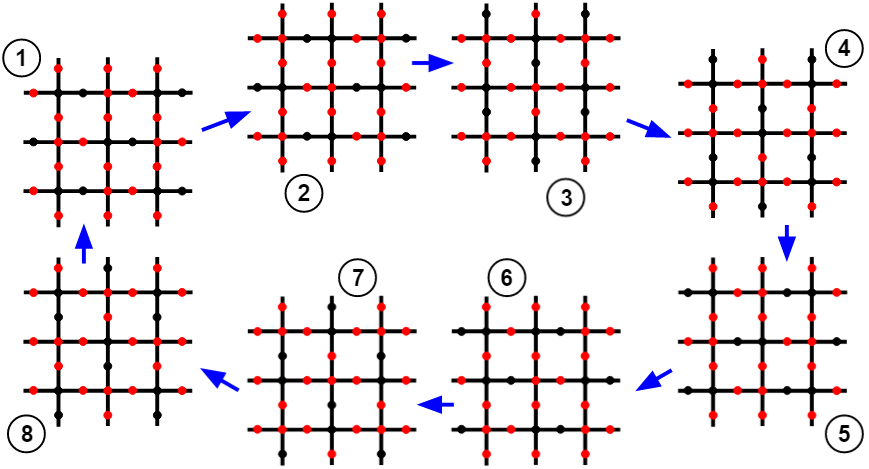}
    \caption{Measurement protocol.  Red vertices indicate the set of repeatedly measured sites, while black sites are unmeasured (the free evolving sets, $A_i$). 
    The adjacent pairs of black vertices trace out an inherently chiral (in this case clockwise) path along a decorated square inside the Lieb lattice.  The path can be made counter-clockwise if the order of the 8 steps is reversed.}
    \label{fig:meas_protocol}
\end{figure}

\begin{figure}[h]
   \centering
    \includegraphics[width=0.35\textwidth]{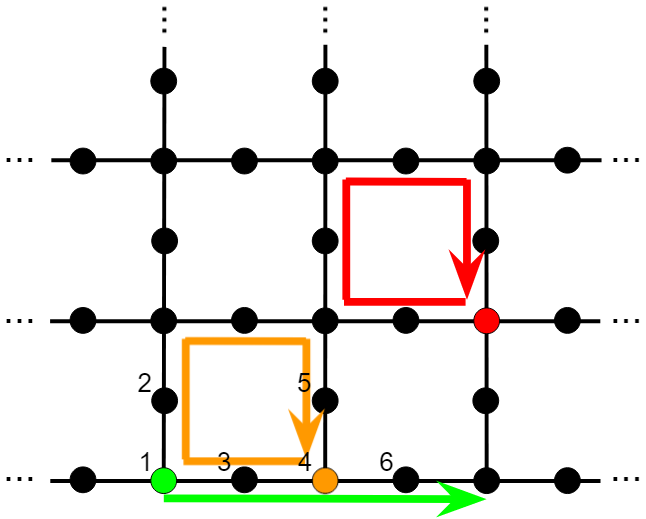}
    \caption{Particle trajectories on the Lieb lattice under the measurement protocol in the infinite measurement (Zeno) limit with the perfect switching cycle ($\frac{T}{8}=\frac{\pi}{2}$).  In this regime, evolution becomes deterministic and particle trajectories can be seen explicitly.  Particles localized in the bulk (Red) at the start of the protocol and particles initialized at sites of type 3 or 4 (orange) on the edge trace out a closed loop after no more than 5 measurement cycles.  On the other hand, particles initialized at sites of type 1 (green) or 6 on the boundary  at the beginning of the protocol propagate along the edge, shifting by 1 dynamical unit cell every 2 measurement cycles (see appendix \ref{appendix deterministic hopping} for details).}
    \label{fig:full loop}
\end{figure}

The measurement cycle consists of 8 steps taking an overall time $T$. At each step, we take repeated snapshots of the presence of particles throughout a subset of the lattice, while the system is allowed to evolve freely between the measurements. We will denote the set of sites {\it not} being measured at step $i$ by $A_i$ as marked in figure \ref{fig:meas_protocol}, and enforce periodicity by setting $A_{i+8} = A_i$.  Within each step, the following procedure is followed:  
\begin{enumerate}
    \item Particle densities at all sites in $(A_i \cap A_{i-1})^c$ are measured.
    \item Free evolution under a free hopping Hamiltonian ${\cal H} = -t_{hop} \sum_{\langle {\bf r}{\bf r'}\rangle }a^{\dag}_{\bf r}a_{\bf r'}$ for a time $\tau=\frac{T }{8 n}$. Here $n$ is an integer describing the measurement frequency. 
    \item Particle densities at all sites in $A^c_i$ are measured.
    \item Steps 2 and 3 are repeated $n$ times. 
\end{enumerate}
For convenience, throughout the paper we will set $t_{hop}=1$ and $\hbar=1$.  For clarity, we note here that in the rest of the manuscript we will refer to one complete iteration of the full 8 step procedure as a ``full measurement cycle'' or sometimes just ``measurement cycle.''  On the other hand, each of the individual steps within the 8 step procedure will be referred to as a ``measurement step.''  

The steps detailed above correspond to a sequence of maps on the density matrix $\rho$ of the system. Two distinct elements are involved in the dynamics. First, the measurement of the presence of a particle at a lattice site ${\bf r}_i$ can be represented by the Krauss map $\rho\rightarrow n_i\rho n_i+(1-n_i)\rho (1-n_i)$, with $n_i=a^{\dag}_i a_i$  the number operator for the site. In between such measurement steps we have unitary evolution, which is described, as usual, via $\rho\rightarrow {\cal U}\rho {\cal U}^{\dag}$, where ${\cal U}$ is a many body evolution operator.

To describe the densities and correlations in the system, we concentrate on the iterative evaluation of two point correlation operators \begin{eqnarray}
     G_{{\bf r}{\bf r'}}(t)=\Tr (\rho(t) a_{\bf r}^\dag a_{\bf r'} ).
\end{eqnarray} 
The two point correlation has a closed evolution equation under particle density measurements and free evolution operations, as long as the free evolution is non-interacting  \cite{klich2019closed}. 
The change in $G$ due to the Krauss map associated with {\it single} site particle density measurement can be shown to imply eliminating correlations between the measured site and other sites \cite{klich2019closed}. Explicitly, one can check that the measurement of particle presence at a lattice site ${\bf r}$ is described by the map
\begin{eqnarray}
G\rightarrow (1-P_{{\bf r}})G (1-P_{{\bf r}})+P_{{\bf r}}GP_{{\bf r}}
\label{particle measurement transformation}
 \end{eqnarray}
where $P_{\mathbf{r}}=|\mathbf{r} \rangle \langle\mathbf{r}|$ is the (single-particle) projector onto site $\mathbf{r}$.  
For non-interacting evolution, fermion operators transform as ${\cal U}^{\dag}a^{\dag}_{\bf q} {\cal U}=U_{\bf q q' }a^\dag_{\bf q'}$, where $U$ is called a single-particle evolution. In this case $G$ transforms as 
    \begin{eqnarray}
 G \rightarrow  U G U^{\dag}.
\end{eqnarray}
In the case of interest for us here, we will take $U=e^{-i \tau H}$, where $H=\sum_{\langle\mathbf{r},\mathbf{r'}\rangle}|\mathbf{r} \rangle \langle\mathbf{r'}|$, describing free hopping of the fermions on the lattice.

To study the repeated application of these maps to $G$, it is convenient to view $G$ as a vector in $\mathcal{H}_{double}=\mathbb{C}^{{\cal N}^2}$ where ${\cal N}$ is the total number of fermion sites. We write $G = \sum_{\mathbf{r}\mathbf{r'}} G_{\mathbf{r}\mathbf{r'}} |\mathbf{r} \rangle \langle \mathbf{r'}| \rightarrow G = \sum_{\mathbf{r}\mathbf{r'}} G_{\mathbf{r}\mathbf{r'}} |\mathbf{r} \rangle \otimes|\mathbf{r'}\rangle$ and the evolution of $G$ under the maps above can be notated as
\begin{eqnarray}
    G(t+T) = \Lambda G(t).
\end{eqnarray}
where $\Lambda$ is the (super) operator acting on $G$ corresponding to the 8 step measurement protocol given in the previous section. To construct $\Lambda$, we write the transformation associated with free evolution and with particle measurement, respectively, as
\begin{eqnarray}&
  G  ~ & \rightarrow \left(U \otimes \Bar{U} \right)G \\   & G ~ & \rightarrow \pi_{\mathbf{r}} G
\end{eqnarray}
where $\pi_{\mathbf{r}} \equiv (1-P_{{\bf r}}) \otimes (1-P_{{\bf r}})+P_{{\bf r}} \otimes P_{{\bf r}}$.  If all the sites in a set $A^c$ are measured simultaneously, we find (See appendix \ref{appendix some derivation details}) that the combined operation on $G$ becomes
\begin{eqnarray}
    \prod_{{\bf r} \in A^c} \pi_{{\bf r}} \equiv \Pi_A = \sum_{{\bf r} \in A^c} P_{\bf r} \otimes P_{\bf r} + P_A \otimes P_A 
    \label{productOfprojections}
\end{eqnarray}
where 
\begin{gather}
    P_A \equiv \sum_{{\bf r} \in A} P_{\bf r} .\label{defPA}
\end{gather}  
Note that $ (\Pi_A G)_{\mathbf{r r'}}=G_{\mathbf{r r'}}$ if both sites $\mathbf{r r'}$ are in the unmeasured set $A$, on the other hand if $\mathbf{r}$ or $\mathbf{r'}$ are in $A^c$ we have $ (\Pi_A G)_{\mathbf{r r'}}\delta_{\mathbf{r r'}}$. In other words, the correlations between the measured sites $A^c$ and all other sites are destroyed while acting as an identity on the subspace $A$ of unmeasured sites. It is important to note that $\Pi_A$ is itself a projection operator on $\mathcal{H}_{double}$. To see this, note that the operators $\pi_{\mathbf{r}}$ form a set of commuting orthogonal projectors, and consequently their product is an orthogonal projector. Another useful property that follows is that 

\begin{gather}
    \Pi_B \Pi_A=\Pi_{A \cap B} .
\end{gather}

We are now in position to write the evolution operator $\Lambda$ describing a cycle of measurements and evolution as described by the measurement protocol above.  Explicitly, after each cycle, which involves $8$ steps each repeated $n$ times, $G\rightarrow \Lambda G$ with 
\begin{eqnarray}
     & \Lambda = \left[ \Pi_{A_8} (U \otimes \Bar{U}) \Pi_{A_8} \right]^n \label{eqn: Lambda}\\
    &  \times \left[ \Pi_{A_7} (U \otimes \Bar{U}) \Pi_{A_7} \right]^{n}...\left[ \Pi_{A_1}  (U \otimes \Bar{U}) \Pi_{A_1} \right]^n \nonumber
\end{eqnarray}
We now turn to analyze the dynamics described by this operator.


\section{The Zeno limit}

We first study the operator $\Lambda$, of Eq.  \eqref{eqn: Lambda}, in the limit of many measurements per cycle (i.e. $n\rightarrow \infty$). The dynamics under high frequency repeated measurements is known as the quantum Zeno limit. The signature characteristic of this regime is the freezing of evolution in the subspace of measured sites.  The Zeno effect (and the closely related anti-Zeno effect) has a long history \cite{Facchi_2008} with broad applications including, for example, counterfactual quantum computing and communication \cite{Hosten2006,Cao2017} and loss suppression in ultracold molecule experiments with strong, long-range dipolar interactions \cite{Yan2013,Zhu2014LossSuppression}.  Over the past 30 years, the Zeno and related effects have been observed experimentally across a variety of physical systems \cite{Itano1990Zeno, Fischer2001Zeno,Streed2006Zeno,Kilina2013ZenoDot,Schafer2014ZenoDyn,Signoles2014Zeno,Harrington2017Zeno}.

Let us first consider one of the eight steps in $\eqref{eqn: Lambda}$. Formally expanding in $\tau=\frac{T }{8 n}$,we find that:
\begin{gather}
     \left(\Pi_{A} (U \otimes \Bar{U}) \Pi_{A} \right)^n = \Pi_A   \left( U_A^n \otimes \Bar{U}_A^n \right) \Pi_A + O(\tau^2 n) \label{eqn: A(U prod U)A terms}.
\end{gather}
Here $U=e^{-i \tau H}$ and $U_A=e^{-i \tau H_A}$ where $H_A \equiv P_A H P_A$. 
To get Eq. \eqref{eqn: A(U prod U)A terms}, we first expand each measurement/evolution step in $\tau$:
\begin{eqnarray}
     & \Pi_{A} (U \otimes \Bar{U}) \Pi_{A} = \Pi_A (I - i \tau  \left[H \otimes I - I \otimes H \right] )\Pi_{A} + O(\tau^2) \nonumber \\ 
     & =\Pi_{A} e^{-i \tau \Pi_A \left[H \otimes I - I \otimes H \right] \Pi_A} \Pi_{A}+ O(\tau^2) \label{eq: one meas step}.
\end{eqnarray}
A short calculation (see appendix \ref{appendix some derivation details}) shows that
\begin{eqnarray}
 \Pi_{A} \left[H \otimes I - I \otimes H \right] \Pi_{A} = H_A \otimes P_A - P_A \otimes H_A \label{eq: a relation}
\end{eqnarray}
hence
\begin{eqnarray}
     & \Pi_{A} (U \otimes \Bar{U}) \Pi_{A} = \Pi_{A} (U_A \otimes \Bar{U}_A) \Pi_{A}+ O(\tau^2)
\end{eqnarray}
which, using $[U_A \otimes \Bar{U}_A,\Pi_A]=0$,  gives \eqref{eqn: A(U prod U)A terms}. The expression \eqref{eqn: A(U prod U)A terms} shows that, in the Zeno limit, the average evolution is dominated by local evolution in the region $A$ and suppresses hopping into the measured sites $A^c$.
Finally, plugging Eq. \eqref{eqn: A(U prod U)A terms} into Eq. \eqref{eqn: Lambda}, we find
\begin{eqnarray}
     &\Lambda = \Pi_{A_8}\left( U_{A_8}^n \otimes \Bar{U}_{A_8}^n \right) \Pi_{A_8 \cap A_7} \label{eq:lambda final} \\
     & \times \left( U_{A_7}^n \otimes \Bar{U}_{A_7}^n \right)...\Pi_{A_2 \cap A_1} \left( U_{A_1}^n \otimes \Bar{U}_{A_1}^n \right)\Pi_{A_1} + O(\tau^2 n). \nonumber  
\end{eqnarray}
Next, we use this result to formally describe evolution for $N$ cycles, when $N n\tau^2\ll 1$, and $n\tau$ is kept constant. 




\begin{figure}
    \centering
    \includegraphics[width = 0.4\textwidth]{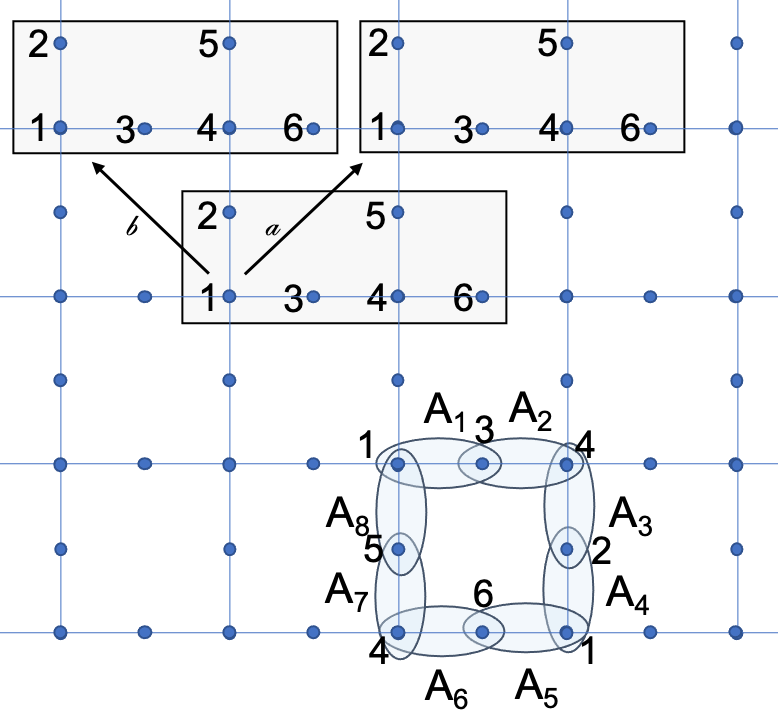}
    \caption{The unit cell for the measurement driven Lieb lattice dynamics consists of two Lieb lattice unit cells. A choice for such a dynamical unit cell is depicted. }
    \label{fig:dynamicalCell}
\end{figure}

\section{Stochastic description of the Zeno limit}

The local nature of the evolution in the Zeno limit \eqref{eq:lambda final} leads to a striking simplification that we now describe. We observe that, if one only follows the local particle density given by the diagonal elements $G_{\bf r r}$, the evolution is given by a periodic classical stochastic process.
To see this, note that the evolution of $G$ in the Zeno limit consists of steps of the form 
\begin{eqnarray}
     G\to \text{  }U_{A_i}^{n}G
   U_{A_i}^{n  \dagger}\label{local_evo}
\end{eqnarray}
Each set ${A_i}$ consists of the union of pairs of neighbouring sites,  the black sites in Fig. \ref{fig:meas_protocol}. Since the pairs forming $A_i$ are disjoint, the evolution $U_{A_i}$, can only develop non-trivial correlations between the sites of the same pair. Consider a pair of such sites. After the evolution, all sites are measured except for sites in ${A_i \cap A_{i+1}}$, which is a set of isolated points on the lattice, in particular, any correlations (non diagonal terms in $G$) developed between the pair of sites in $A_i$ would be set to zero once the site in $A_i$ but not in $A_{i+1}$ is measured (long range correlations between sites in ${A_i \cap A_{i+1}}$ are not annihilated, but will be annihilated in the next step, and cannot be generated by any of the $U_{A_i}$). Thus, if we start with a diagonal $G$ it will remain diagonal throughout the evolution. Moreover, even if we start with some non-zero off-diagonal terms, these will be quickly annihilated by the measurements. Thus we should be able to describe the evolution, in the Zeno limit, just in terms of the dynamics of the diagonal of $G$. Indeed, note that $G_{\bf r r}$ are real non-negative numbers and the total number of particles $\sum_{\bf r}G_{\bf r r}$ is a constant of motion, and thus $G_{\bf r r}$ can be treated (up to normalization) as probabilities, and the process describing the evolution is a  classic stochastic process. 


Explicitly, if we represent the density at time $t$ as a vector, $\ket{g(t)}$, defined via
\begin{eqnarray}
[g(t)]_{\bf r}\equiv
   G_{\bf r r}(t),
\end{eqnarray}
then in the Zeno limit the density evolves via Markovian dynamics as
\begin{eqnarray}  \ket{{g}}\longrightarrow  R_{cyc} \ket{{g}}   \end{eqnarray}
   where the transition matrix $R_{cyc}$ consists of the 8 steps of our process, namely \begin{eqnarray}
        R_{cyc}=R_{8}R_{7}R_{6}R_{5}R_{4}R_{3}R_{2}R_{1}. \label{Rproduct}
   \end{eqnarray}
The transition matrices $R_{i}$ are defined as follows.  Each unmeasured set $A_i$ is associated with two site types $\alpha$, $\beta$ that are not being measured, as described in Fig. \ref{fig:dynamicalCell} (e.g. $A_1$ is the union of all sites of types $1,3$; $A_2$ is the union of sites $3,4$; etc). The unitary evolution associated with a given unmeasured set $A_i$, breaks into a sum of pairs of nearest neighbours: \begin{eqnarray}\label{evAi}
     U_{A_i}=I_{A_i^c} \bigoplus_{\langle\alpha,\beta\rangle\in  A_i} e^{i \tau n \sigma_x}.
   \end{eqnarray} 
   Next, we apply the evolution  \eqref{local_evo} and then measure all sites except those in $A_i\cap A_{i+1}$, which has the effect of eliminating off-diagonal elements in $G$. Consider one of the pairs of sites $\langle\alpha,\beta\rangle\in  A_i$ and an initially diagonal $G=diag(g_1,g_2)$.  Applying the evolution \eqref{evAi} to get $e^{i \tau n \sigma_x} G e^{-i \tau n \sigma_x}$ and then setting the off diagonal elements to zero, we get a new diagonal matrix $G$ with $G=diag(\cos^2({n \tau })g_1+\sin^2({n \tau }) g_2,\cos^2({n \tau })g_2+\sin^2({n \tau })g_1)$. In other words, a particle located in one of the sites jumps to the other site with probability \begin{eqnarray}\label{eq:hopping p}
        p=\sin^2({n \tau })=\sin^2({T \over 8 }),
   \end{eqnarray} or stays with probability $1-p$. A particle in any other position will not move. Therefore:
   \begin{eqnarray}  R_i=\oplus_{\langle\alpha,\beta\rangle\in  A_i}\left(
\begin{array}{cc}\small
 1-p  &  p  \\
  p  & 1-p \\
\end{array}
\right)\oplus
   _{\text{other sites}} I
   \label{eq:R stochastic}
   \end{eqnarray}
   This defines a periodically driven random walk. We note that the transition matrices $R_{i}$ are bi-stochastic matrices, and thus so is $R_{cyc}$. 
   
A remark is in order here about Eq. \eqref{eq:R stochastic}.  In a system with a boundary, a set $A_i$ may include isolated sites that do not have an adjacent neighbour also in $A_i$.   For example consider the boundary of the lattice in Fig. \ref{fig:dynamicalCell}. The set $A_3$ as defined includes sites of type $4$ and $2$, however looking at the lower boundary, we see that sites of type $4$ on the boundary do not have an adjacent site of type $2$. Similarly to the measured sites, the dynamics for these isolated elements of $A_i$ are frozen in the Zeno limit.  In \eqref{eq:R stochastic}, the isolated elements of $A_i$ are included in ``other sites'' since they are not part of an adjacent pair in $A_i$. 
   
   The particular choice $T=4\pi$, leads to $p=1$. We refer to this choice as "perfect switching". In this case, $R_{cyc}$ is a permutation matrix, and the motion of particles is deterministic. Of course, on the other hand, when $T=8\pi$,  $p=0$ and there is no evolution at all.
   
We now consider the counting statistics of transport to the right per cycle. To do so, we attach a counting field $e^{i\theta}$ to each horizontal link, by modifying the above transition matrices of $R_1,R_2,R_5,R_6$ to
 \begin{eqnarray}  R_{i }=\oplus_{\langle\alpha,\beta\rangle\in  A_i}\left(
\begin{array}{cc}
 1-p  & e^{i\theta}p  \\
 e^{-i\theta}p  & 1-p \\
\end{array}
\right)\oplus
   _{\text{other sites}} I   \label{eq:Ri with theta} \end{eqnarray}
   whenever $\alpha,\beta$ are nearest neighbours on a horizontal line, such that $\alpha$ is to the left of $\beta$.
   
With the counting field present, we can introduce the moment generating function, 
  \begin{eqnarray}&
   \chi_{N}(\theta) \equiv \sum_{ij} \sum_{w:i\rightarrow j} e^{i \theta {l(w)}} Prob_N(w)G_{ii}(0) \nonumber \\ &=  \sum_{ij} [R_{cyc}(\theta)^N]_{ij}   G_{ii}(0)=  \langle
 I|R_{cyc}^N (\theta)\ket{g_0}
  \end{eqnarray}
where $w:i\rightarrow j$ is a sequence of hops from site $i$ to site $j$, $Prob_N(w)$ is the probability for the path $w$ after $N$ measurement cycles using the transition matrix $R_{cyc}$, and $l(w)$ is the net number of hops in the $x$ direction. In the next line, $|g_0\rangle$ is the initial density distribution at $t=0$ and $|I\rangle$ is a vector whose elements are all $1$ (corresponding to $G=I$).
   
We can use $\chi_N$ to compute quantities of interest, most important of which is the flow, defined as the total displacement per cycle, per unit length. The flow in the $x$ direction is given by 
\begin{eqnarray}
   F =\lim_{N\rightarrow \infty}F_N
\end{eqnarray}
where $F_N$ is the average flow in the first $N$ cycles,
\begin{eqnarray}
   F_N = {1\over L_x}{1\over N }\left.i \partial _{\theta }  \chi_N(\theta) \right|_{\theta=0} \label{currentformula}
\end{eqnarray}
with $L_x$ the length in the $x$ direction. 
 
\subsection{Absence of bulk transport.}
In a translation invariant situation, it is convenient to work in momentum space. Here, we must use the ``dynamical unit cel'' where the periodic evolution happens, which is double the Lieb lattice's original unit cell (see Fig. \ref{fig:dynamicalCell}). 
   
The Bravais lattice for the dynamical unit cell is a rotated square lattice whose primitive Bravais vectors are marked as $a,b$ in  Fig. \ref{fig:dynamicalCell}. Below, we use $\pmb{n},\pmb{m}$ to denote the position of the unit cell and $\mu,\nu\in \{1,..,6\}$ to denote the individual atom inside the cell. We can then write:
\begin{eqnarray}\label{bulkFourier}
        R_{i}(\pmb{n},\mu;\pmb{m},\nu)=\int \frac{d^2 k}{2\pi} R_{i}(\pmb{k},\mu,\nu)e^{i \pmb{k}\cdot (\pmb{n}-\pmb{m})}.
\end{eqnarray} for example, $R_5$, is associated with $A_5$, which includes sites $1,6$ in neighbouring dynamic unit cells, hence
\begin{eqnarray*}\small
R_{5}(k,\theta)=\left(
\begin{array}{cccccc}
 1-p & 0 & 0  & 0 & 0 & p e^{-i(\theta+k\cdot(a-b))} \\
 0 & 1 & 0 & 0 & 0 & 0 \\
 0 & 0 & 1 & 0 & 0 & 0 \\
 0 & 0 & 0 & 1 & 0 & 0 \\
 0 & 0 & 0 & 0 & 1 & 0 \\
 p e^{i(\theta+k\cdot(a-b))} & 0 & 0
   & 0 & 0 & 1-p 
   \label{eq: R1k}
\end{array}
\right).
   \end{eqnarray*}
In the deterministic case, $p=1$, we find for the full cycle
   \begin{eqnarray}\small
R_{cyc}(k,\theta)=\left(
\begin{array}{cccccc}
 1 & 0 & 0 & 0 & 0 & 0 \\
 0 & 0 & e^{i\theta} & 0 & 0 & 0 \\
 0 & 0 & 0 & e^{i\theta} & 0 & 0 \\
 0 & 0 & 0 & 0 & 1 & 0 \\
 0 & 0 & 0 & 0 & 0 & e^{i k\cdot b} e^{-i\theta} \\
 0 & e^{-i k\cdot b} e^{-i\theta} & 0
   & 0 & 0 & 0 
   \label{eq: Rcycle}
\end{array}
\right).
   \end{eqnarray}
It is possible to check that in this case, with $p=1$, $R_{cyc}(k,\theta)^5=I$. Therefore, the system returns to itself after 5 cycles without generating any transport at all.
For $p\neq 1$, we find that $Re Tr R_{cyc}(k,\theta)^{n}$ is a symmetric function of $\theta$, and here too, there is no transport after an arbitrary number of cycles. To do so we computed the characteristic polynomial of the matrix $R_{cyc}(k,\theta)$ and found that it is equal to that of $R_{cyc}(-k,-\theta)$, implying equality of eigenvalues of the matrices.

It is also possible to check that for any $k_x,k_y \neq 0 \pmod{2\pi}$, $||R_{cyc}||<1$, which implies the long time behavior will be dominated only by the $k=0$ component of the initial distribution. For $k_x=k_y=0$ and $\theta=0$ there is a single  left and right eigenvector with eigenvalue $1$, which is the uniform density state $|{I} \rangle $, implying that up to exponentially small corrections, the current density \eqref{currentformula} vanishes.

\subsection{Edge transport}
We have concluded that there is no bulk transport associated with the stochastic process defined by $R_{cyc}$, for any $p$. In this section, we contrast the situation to when an edge is present.
We implement the dynamics by removing all sites beyond the physical edges (e.g. sites with $y<1$) and removing any transitions involving sites beyond the edges from the dynamics.
We start with the deterministic case, namely $p=1$. In Fig. \ref{fig:full loop}, we exhibit a half plane with an edge. For $p=1$, we can track the motion of each particle and conclude that bulk particles perform a closed loop. On the other hand, particles starting at the edge divide into two sets:  some of the edge particles ($6,1$) perform a motion along the edge, while some ($3,4$) perform a closed loop. Thus, if we start from an initial state where particles are placed along the edge we will have particle transport along the edge (particles $6,1$ will move to the right). This behavior is clearly analogous to the familiar skipping orbits in the semi-classical description of the integer quantum Hall effect.


What will happen away from $p=1$? Consider first the case of a strip with periodic boundary conditions in the long direction (say $x$) and open boundary conditions in the $y$ direction with $L_y$ dynamical unit cells in the $y$ direction. Let us consider states that are translationally invariant in the $x$ direction, allowing us to analyze the behavior in \eqref{currentformula} in momentum space. For any momentum $k_x$, the transition operator $R$ can then be written as a $6L_y\times 6L_y$ matrix and analyzed. For $0<p<1$, any initially positioned particle has a finite probability to get to any other site within a finite time and the only steady state distribution of $R$ with eigenvalue $1$ is that of uniform density (in contrast to the $p=1$, where additional steady states are possible. 
This distribution will be approached exponentially fast, governed by ${\lambda_2}^N$ where $\lambda_2$ is the second largest eigenvalue of $R$. In the uniform density distribution, there is no net charge transfer. Indeed in that case, the charge transfer of the upper and lower edge is carried in opposite directions and cancels. 

To get net transfer, we initially place particles only close to one of the edges. In a finite width system, away from the perfect switching cycle, we expect the charge transport to be transient: once the measuring protocol starts, it will transport a finite amount of particles while also spreading particles towards the second edge, rapidly approaching the uniform density state. Thus, to study the net particle flow associated with a given edge we must work in the thermodynamic limit ($L_y \rightarrow \infty$), or, more precisely, $L_y\gg T_w$ where $T_w$ is the typical time it may take a particle to diffuse from the middle of the sample to one of the edges. 

\begin{figure}[h]
    \centering
    \includegraphics[width=0.35\textwidth]{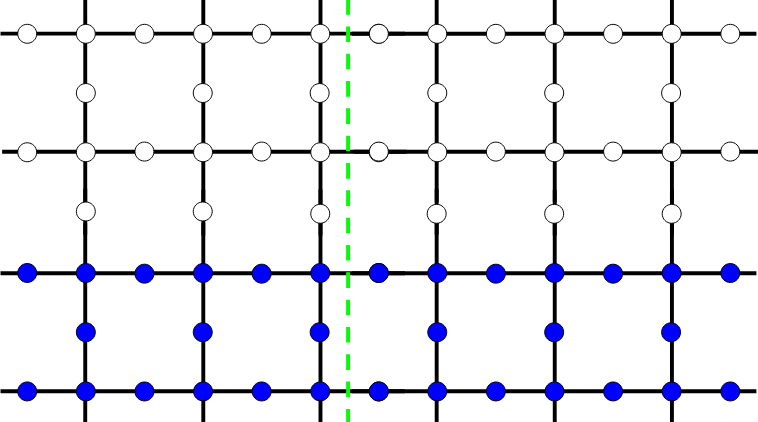}
    \caption{Lieb lattice with lower-half plane filled with particles (in blue).  Trace is taken of the half plane to the right of the green dashed line.  The flow across the barrier is then given by the difference between the right-half trace before and after evolution.}
    \label{fig:Flow Diagram}
\end{figure}

We now numerically compute the number of particles, $F$, that flow across a slice through the Lieb lattice during evolution (see figure \ref{fig:Flow Diagram}). In other words, we compare the number of particles to the left of the slice before and after the application of $\Lambda$, computing:
\begin{equation}
    F_{sim} \equiv \sum_{ \text{{\bf r} to the left of slice}}\big((\Lambda G)_{\bf r r}- G_{\bf r r} \big).
    \label{eq: Fsim}
\end{equation}
This is done by initiating the system at $G(t=0)=G_0$ where $G_0$ is a diagonal matrix corresponding to particles placed on the bottom half of a square lattice, with open boundary conditions. We then iterate the map \eqref{eqn: Lambda}, computing $\Lambda^N G_0$, increasing $N$ but being careful to limit the number of cycles to remain within the regime that no significant density has had time to build up close to the upper edge.

\begin{figure}[h]
    \centering
    \includegraphics[width=0.35\textwidth]{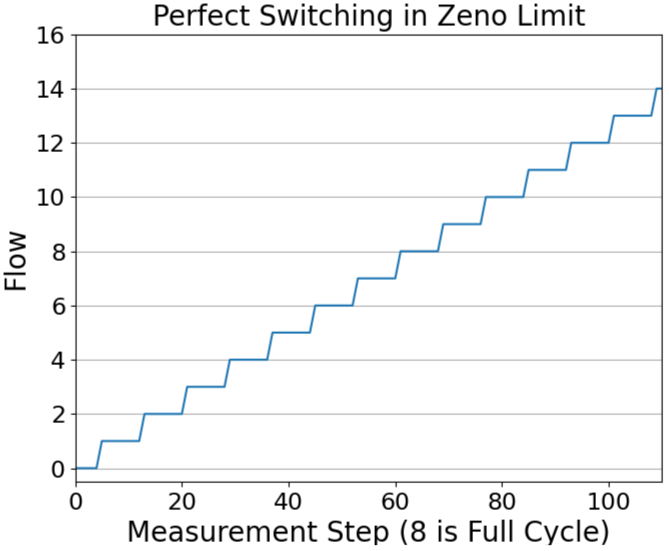}
    \caption{Charge transfer of the left-half filled plane in the Zeno limit with $\frac{T  }{8} = \frac{\pi}{2}$, namely $p=1$.  In this section, the Lieb lattice size for all simulations is $33\times33$  unless otherwise stated.  Here, precisely one particle is transported across the flow cut during the 8 step  measurement cycle.}
    \label{fig:Zeno Flow}
\end{figure}

The combination of the Zeno limit with the perfect switching point $p=1$ leads to a clearly quantized flow, as is clearly exhibited in  Fig \ref{fig:Zeno Flow}, and can be understood by tracking the trajectories of the particles (see Fig.  \ref{fig:full loop} and appendix \ref{appendix deterministic hopping} for details of the motion). 
Next we will consider both the cases of $p\neq 1$ as well as the non-Zeno limit.

\section{Charge Transport: Bulk-Edge Decomposition} 
\begin{figure}
    \centering
    \includegraphics[width=0.5\textwidth]{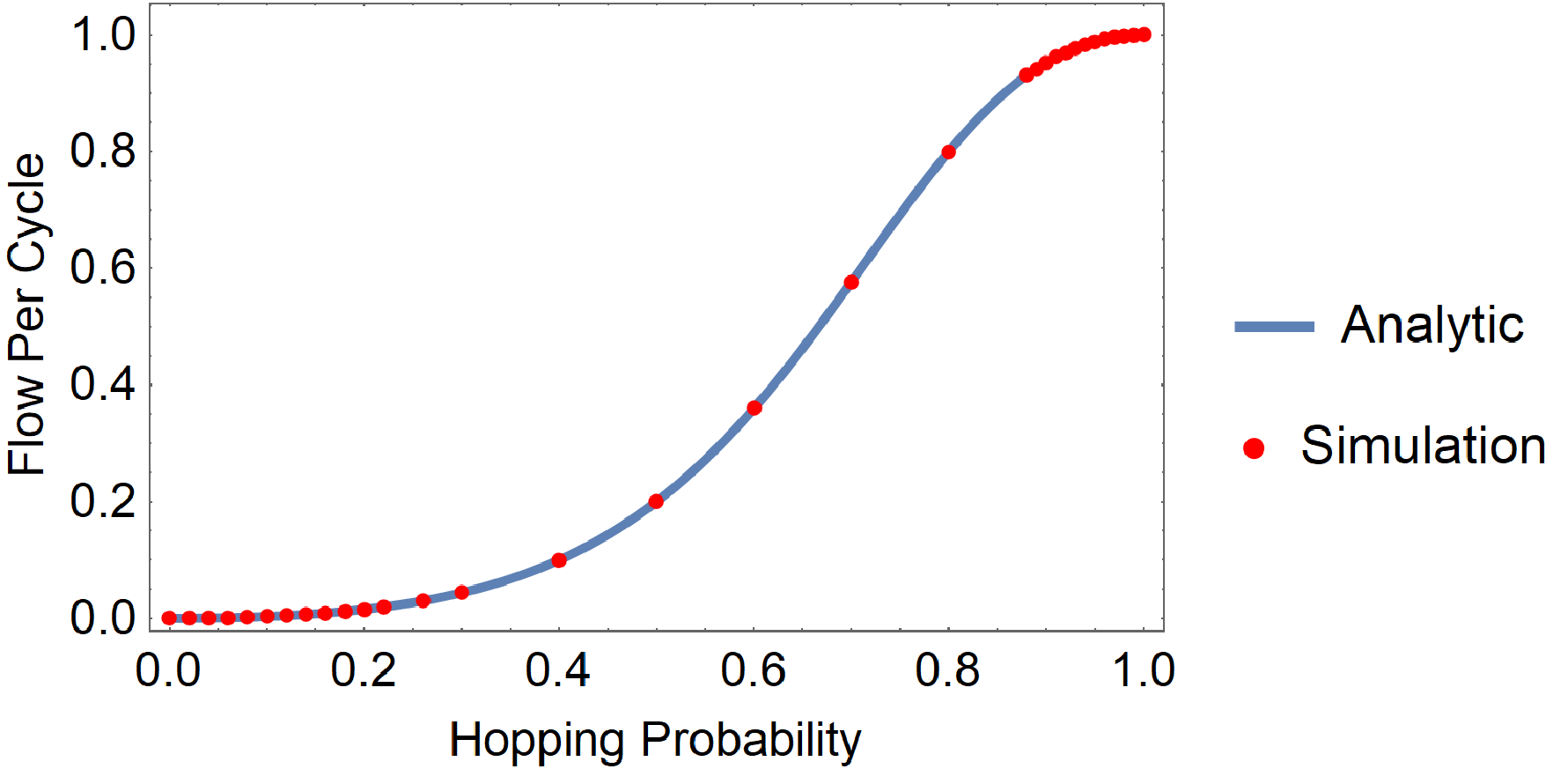}
    \caption{Charge transport per measurement cycle in the Zeno limit.  The analytic formula given in Eq. \eqref{eq:F final} is compared with the transport found from direct simulation for a selection of hopping probabilities.}
    \label{fig:analytic flow per cycle}
\end{figure}

We now turn to calculating the charge transport per measurement cycle in the Zeno limit with arbitrary $p$. The result is described in Fig. \ref{fig:analytic flow per cycle}.  Since the bulk transport vanishes for any $p$, the flow will be still completely localized near the edge. Below, we exhibit an analytical formula for the flow, Eq. \eqref{eq:F final},  achieved using  a bulk-boundary decomposition in the limit $L_y\rightarrow\infty$ (and verify it by direct numerical simulations of the dynamics on finite systems). The resulting dependence on $p$ is shown in Fig. \eqref{fig:analytic flow per cycle}, exhibiting a crossover behavior ranging from the integer transport at $p=1$ to no transport when $p=0$ (where the dynamics is trivial, since all hopping is blocked).

We show how the edge flow can be written in terms of bulk operators.  This correspondence between the bulk properties of the system and the charge transport on the edge provides both a direct, efficient method to calculate the flow as well as implies the robustness of the flow to any perturbations near the boundary of the system.  

\begin{figure}
    \centering
    \includegraphics[width=0.45\textwidth]{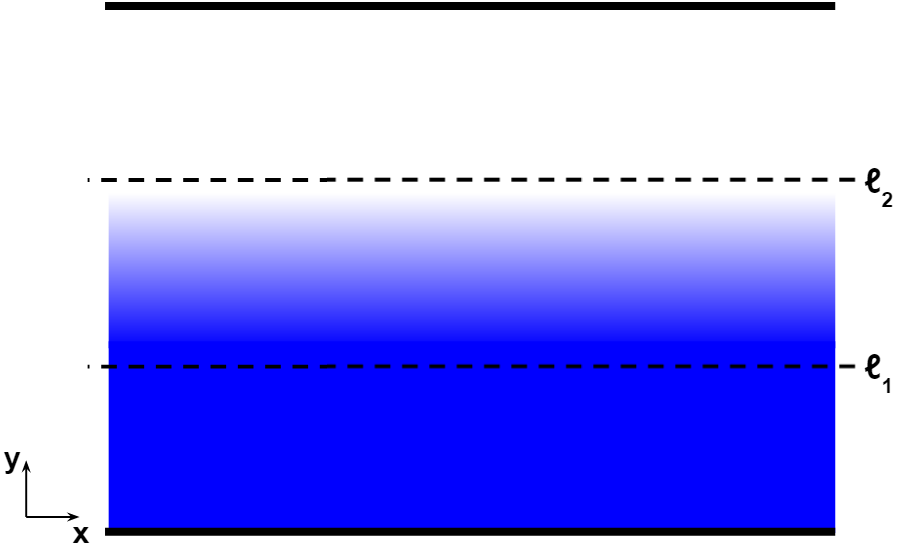}
    \caption{Initial particle density chosen for the flow  analysis.  All sites below the line $y=\ell_1$ are filled with particles (shown in blue).  All sites above $y=\ell_2$ are empty.  The probability of finding a particle at sites in-between $y=\ell_1$ and $\ell_2$ is left arbitrary as the charge density in this region will not affect the flow.}
    \label{fig:Gm description}
\end{figure}

To observe the flow we imagine an infinite strip in the $x$ direction. We partition the strip into 3 regions as shown in Fig. \ref{fig:Gm description}.  The bottom region of the system (below height $\ell_1$) is completely filled with particles, while the top (above $\ell_2$) is empty.  In-between $\ell_1$ and $\ell_2$, the particle density is left arbitrary and will have no effect on the particle transport. This choice isolates the flow along just the bottom edge of the system, removing the equal and opposite contribution from the flow along the top edge.  Charge distributions of this type are analogously used as a tool to calculate charge flow along an edge in the context of Floquet topological insulators.  See, for instance, \cite{Lindner2016ChargePump}.  In appendix \ref{appendix bulk-edge} we prove that
\begin{eqnarray}
        F = F_{bulk} + F_{edge}  \label{eq:F final}
\end{eqnarray}
where
\begin{gather}
     F_{bulk} = i\sum_{\alpha\beta} [ J_B (\mathbf{k}) \frac{1}{I-R_B (\mathbf{k})}  \partial_{k_y} R_B (\mathbf{k})]_{\alpha\beta} |_{\mathbf{k}=0}\label{eq:F final bulk}
\end{gather}
and
\begin{gather}\label{edgeflow}
F_{edge}=   {1\over L_x}\langle\mathbf{I}|P_{y\leq 3} J P_{y\leq 2} |\mathbf{I}\rangle .
\end{gather}
Here, $R_B$ is a bulk transition operator, equal to $R_{cyc}$ except with periodic instead of open boundary conditions to make it translational invariant. The transition operators $R_B,R_{cyc}$ are used to define appropriate currents $J= -i \partial_{\theta} R_{cyc}(\theta) |_{\theta = 0}$ and similarly $J_B= -i \partial_{\theta} R_{B}(\theta) |_{\theta = 0}$. 
Above, for an operator $A$, translational invariant in $x$ and $y$ with respect to the unit cell of the dynamics and with matrix elements $A_{\alpha\beta}({\bf r},{\bf r}')$, we define $A({\bf k})_{\alpha\beta}$ as in \eqref{bulkFourier}.
In the edge contribution, $P_{y\leq 2}$ is a projection operator on sites with $y\leq 2$. The above expressions are proven starting from the expression Eq. \eqref{currentformula} for the flow $F_N$ after a finite number of cycles and then taking the limit of large $N$ while maintaining $N\ll \ell_1$ and keeping $\ell_2 - \ell_1$ constant.

To compute $F_{edge}$ we can write, explicitly
\begin{gather}
F_{edge}=   {1\over L_x}\sum_{\alpha\beta}\sum_{x,x'=1}^{L_x}\sum_{y'=1}^{2}\sum_{y=1}^{3} J_{\alpha\beta}(x,y;x',y').\label{edge current explicit}
\end{gather}
Calculating $F_{edge}$ in this form we find with our measurement protocol 
 \begin{eqnarray}
      F_{edge} = p^2 + p^3 + p^4.
 \end{eqnarray} 

The contribution of $F_{bulk}$ to the transport can also be evaluated readily, as it is made up of products and an inverse of  $6\times 6$ matrices and so can be easily computed for any $p$. In Fig \ref{fig:analytic flow per cycle}, we combine these two terms and compare with direct simulations of the dynamics which exhibit excellent agreement.

We now make the following especially important remark that \textit{both} $F_{\text{bulk}}$ and $F_{\text{edge}}$ depend only on the bulk properties of the system (assuming weak constraints to be described below).  This implies that the flow is completely insensitive to the details of the structure of the edge or local perturbations.  This can be argued in the following way. 
We first note that
\begin{eqnarray}\label{eq:no_current_in_filled}
     \langle\mathbf{I}|J  |\mathbf{I}\rangle=0
\end{eqnarray}
i.e. there is zero total current in a uniform density system. Eq \eqref{eq:no_current_in_filled} can be seen from the form of the dynamics generated by our $R$ matrices,  
\eqref{eq:Ri with theta}, since  
\begin{eqnarray*}
    -i  \partial_{\theta}|_{\theta=0} R_{i }|\mathbf{I}\rangle=\oplus_{\langle\alpha,\beta\rangle\in  A_i}\left(
\begin{array}{cc}
 0  & p  \\
 -p  & 0 \\
\end{array}
\right)\oplus
   _{\text{other sites}} I   |\mathbf{I}\rangle=0.
\end{eqnarray*}
Now, consider a modification of the stochastic dynamics along the bottom edge of the system still obeying the no total current condition \eqref{eq:no_current_in_filled}, and that there is no explicit bulk current introduced (the latter restriction of no added bulk currents may be removed upon closer analysis, see appendix \ref{appendix: Robustness of Flow}). Assuming the current operator is short ranged (with range of at most one unit cell), one can rewrite the expression \eqref{edgeflow} as 
\begin{gather}\label{edgeflow1}
F_{edge}=   {1\over L_x}\langle\mathbf{I}|J P_{y\leq 2} |\mathbf{I}\rangle = -{1\over L_x}\langle\mathbf{I}|J P_{y> 2} |\mathbf{I}\rangle
\end{gather}
the expression on the right hand side for $F_{\text{edge}}$ is independent of how we vary $J$ on the lower boundary. In other words, the global zero current condition together with the fact that the two edges responsible for the transport are physically separated means that the contribution of $F_{\text{edge}}$ to the flow must be protected. Together with the bulk nature of $F_{\text{bulk}}$, we see indeed a protected flow. A more detailed proof can be constructed as follows (with details in the appendix): Consider the following matrix, describing the perturbed dynamics:
\begin{gather}
    {R}_M (\mathbf{r},\mathbf{r'}) = 
    \begin{cases}
         \tilde{R} & r_y,r'_y \leq \ell_1 - (N+1)\\
         R_{cyc} & \ell_1 -  (N+1) \leq r_y,r'_y \leq \ell_2 + (N+1)\\
         \tilde{R}' & r_y,r'_y \geq \ell_2 +  (N+1) \\
         0 & \text{otherwise}
    \end{cases}
\end{gather}
where $\tilde{R}, \tilde{R}'$ are real matrices such that ${R}_M$ is doubly stochastic, i.e. ${R}_M$ is identical to $R_{cyc}$ in the bulk but modified near the boundary. 

In appendix \ref{appendix: Robustness of Flow}, we prove that the flow of $ {R}_M$ is equivalent to the flow of $R_{cyc}$ assuming that 
\begin{eqnarray} &
    \bra{\mathbf{I}}  {R}_M(\theta=0) = \bra{\mathbf{I}} \hbox{ and }  {R}_M(\theta=0)  \ket{\mathbf{I}} = \ket{\mathbf{I}} \label{eq: cond 1} \\ &
    \langle
   \mathbf{I}| {J}_M |\mathbf{I}\rangle = 0,\label{eq: cond 2}
\end{eqnarray}
where $J_M$ is the current operator associated with $R_M$. The first condition requires that ${R}_M$ preserves particle number and that a uniform density is a steady state of the evolution; this implies that the transition matrix remains doubly stochastic.  The second condition is the requirement that no net current can flow in the completely filled system.  Note, the conditions \eqref{eq: cond 1} and \eqref{eq: cond 2} are certainly satisfied whenever ${R}_M$ is a product of symmetric, doubly stochastic matrices which encapsulates a large class of physically relevant perturbations including, for example, local potentials, local variations
of the hopping parameter, and removal of sites from the
lattice. Indeed, repeating the argument leading to \eqref{eq:hopping p}, including the presence of local potential terms (or variation in $t_{hop}$) in the local Hamiltonian will just locally change the hopping probability $p$, retaining the form of the dynamics as in \eqref{eq:R stochastic} with modified $p$s (i.e. still made of doubly stochastic building blocks). Removal of sites can similarly be described by taking $p=0$ for transitions to the removed site.
Due to its stability, the flow may be viewed as a continuous topological invariant for the system.  We emphasize that such protection cannot be achieved in 1D systems, which can be easily disconnected by the removal of a few sites.       


A technical remark is in order here. The simulation result in Fig \ref{fig:analytic flow per cycle} was computed using Eq  \eqref{eq: Fsim} with the cut defined as shown in Fig. \ref{fig:Zeno Flow}.  Therefore, the quantity computed in the simulations, $F_{sim}$ (Eq.~\eqref{eq: Fsim}), is equivalent to placing the counting field $\theta$ only at a subset of the horizontal edges as opposed to placing $\theta$ on all horizontal edges as was used in defining $R_{cyc}(\theta)$ through Eq. \eqref{eq:Ri with theta}. Accounting for the number of edges included in the simulations - these include 2 edges per two dynamical unit cells - and that each dynamical unit cell involves $4$ edges and  that no charge accumulation occurs, we find simply 
\begin{equation}
 F_{sim} =  \frac{F}{4} 
    \label{eq: flow compare}
\end{equation}
which was used in the comparison Fig. \ref{fig:analytic flow per cycle}.

At this point, we wish to further discuss and clarify the nature of the protection of the flow in \eqref{eq:F final} and in what sense it is localized on the edge.  In our set up, the charge density is constant in a thick  neighbourhood of the edge. 
 It is important to emphasize, however, that the protection is not simply due to Pauli blocking, but a feature of the classical stochastic dynamics. This is evident when we consider the flow when the density in the occupied (blue) region in Fig. 7 is uniformly reduced to a lower density $\rho<1$. In this case (especially at low density), Pauli blocking is not important for the dynamics.  However, the linearity of our stochastic dynamics shows that the new Flow will be $F(\rho)=\rho  F(\rho=1)$. Thus, the flow is protected (in the sense explained above) for any filling $\rho$, in sharp contrast with most topological insulators. 
 
Another interesting feature of the charge transport here is that the flow we compute (for $p\neq 1$) is the result of the collective contribution of fermions that approach the edge, travel along it for a time, and then diffuse away, rather than the result of single wave packets traveling along the edge without dispersing. An alternative perspective that can help clarify the edge nature of the flow can be obtained by adding a particle sink/source where holes/particles can be injected/extracted from the system. In this case holes injected in the bulk will only contribute to charge flow (for a finite time) when they reach the edge. 
Note, the edge flow is due to unbound charges which are only a partial contribution to the local currents in the system.  For example, in the completely filled system, since the density is uniform and the $R$ matrices are symmetric there can be no current on any link in the system.  The net zero current is the result of two different cancellations in the bulk and on the edges of the system. In the bulk, the zero current is the result of local current loops that give rise to a uniform magnetization and the net current is $\del\times M=0$. On the edge, the net current is zero as a result of cancellation between the bound currents, as in the bulk, and unbound currents that exist only on the edge.
The distinction, however, between charge transport (which is localized on the edge) and current (which is not localized on the edge) is largely independent of the present work and similar distinctions must be made, for example, in discussions of Floquet topological insulators \cite{Lindner2016ChargePump}.
 



For topological insulators, a bulk gap implies that small alterations to the bulk Hamiltonian will not destroy an edge mode so long as symmetries protecting the topological phase are preserved \cite{Ando2013TIReview}. The actual value of the current will depend on the density and on the details of how the bands are filled. Similarly, here, small changes in the carrier density will lead to changes in the magnitude of the flow, but not its existence.  Interestingly, unlike topological insulators, the existence of the flow and the protection we discuss are independent of the initial filling, which manifests itself in the off-diagonal part of $G$ when the process starts. 

On the other hand, while here the flow is robust (in the sense explained above) at any density, it's value is not in general robust to arbitrary global changes of the parameters.
In our system, it is possible to continuously change the flow by small extensive perturbations, say, changing the total period $T$.  However, as stated, the exact value for the flow of the system during $N$ cycles is protected against even strong perturbations as long as these are far enough (i.e. within a distance at least $N$) from the interface with the region which is not of uniform density (see Appendix \ref{appendix: Robustness of Flow}). Perturbations within the interface region may alter total charge transport values by inducing bulk currents in the system (see Fig. \ref{fig:flow stability}).

\begin{figure}
\centering
\begin{subfigure}{0.5\textwidth}
    \centering
    \includegraphics[width=\textwidth]{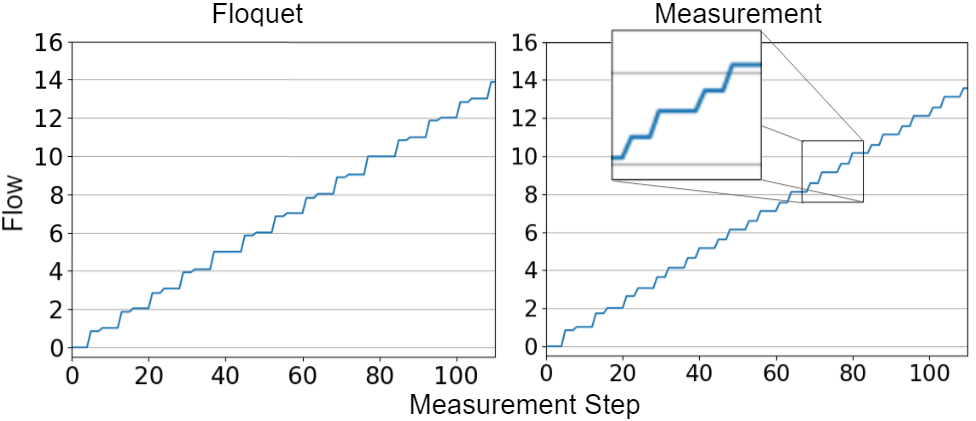}
    \caption{}
    \label{fig:Flow off pi2 compare}
\end{subfigure}
\begin{subfigure}{0.5\textwidth}
    \centering
    \includegraphics[width=\textwidth]{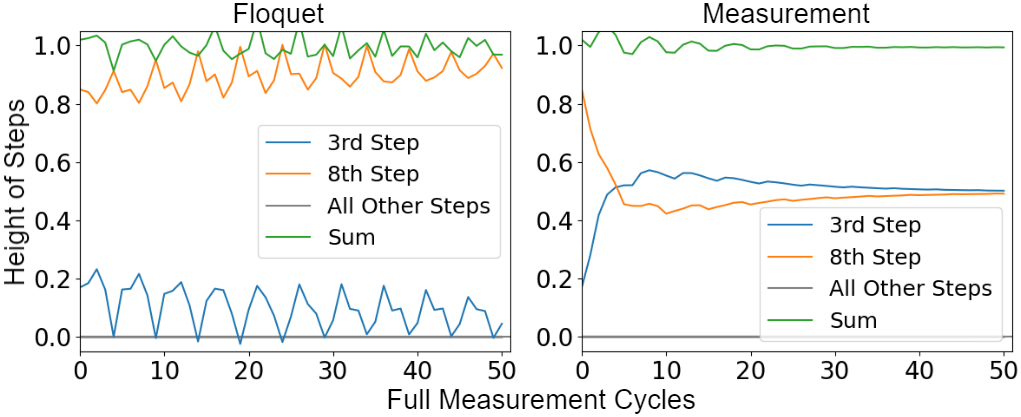}
    \caption{}
    \label{fig:Converge Flow Floquet Compare}
\end{subfigure}
\caption{(a) Charge transfer for Floquet system (left) and measurement protocol in the Zeno limit (right) where, in both cases, the hopping probability $p = 0.96$. (b) Charge transfer after each measurement step for the Floquet system and measurement protocol with hopping probability $p = 0.96$.  Note the convergence of the 3rd and 8th step to half the total flow per cycle.}
\label{fig: compare floquet and meas}
\end{figure}

\begin{figure}
    \centering
    \includegraphics[width=0.5\textwidth]{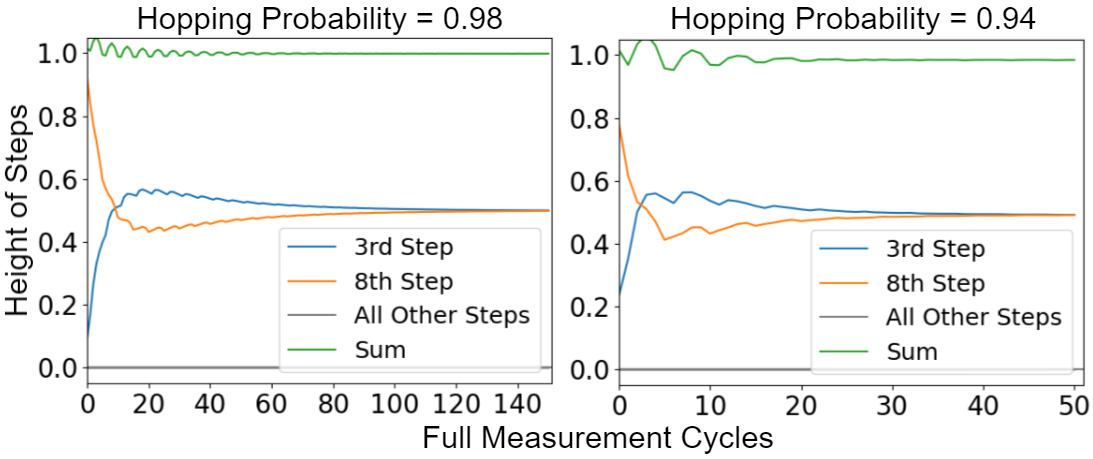}
    \caption{Flow after each measurement step for hopping probabilities $p = 0.98$ (left) and $p = 0.94$ (right).  Note, in the long time limit, the 3rd and 8th measurement step of both hopping probabilities converge to half the total flow per measurement cycle.  }
    \label{fig:Converging flow compare offset}
\end{figure} 

\begin{figure}
    \centering
    \includegraphics[width=0.5\textwidth]{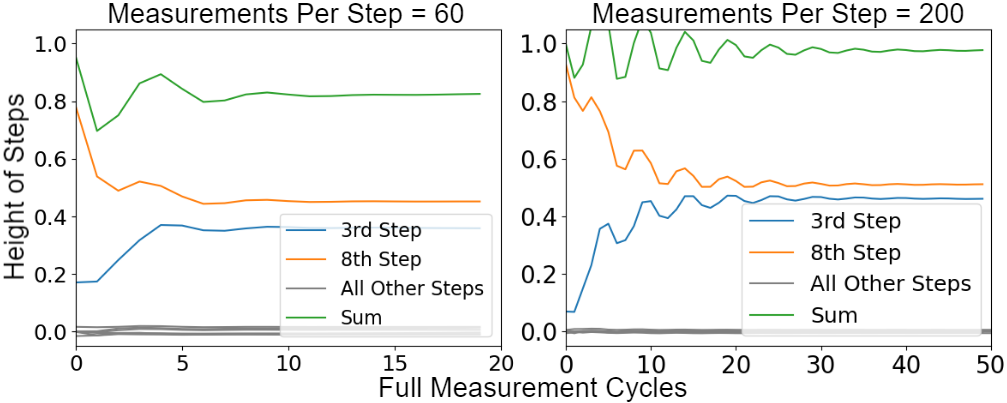}
    \caption{Flow after each measurement step for $n = 60$ and $n = 200$.  For both, $\frac{T  }{8} = \frac{\pi}{2}$.}
    \label{fig:After3 After5 Full Measure Compare}
\end{figure} 


It is interesting to compare the behavior in the Zeno limit with a Floquet topological insulator evolution in our system which is equivalent to the one introduced in \cite{rudner2013anomalous}. There, a periodic driving protocol is used as the source of chirality in the system, where hopping between neighbouring sites are sequentially turned on, but without any measurements.  Explicitly, the analogous evolution for us, $\Lambda_{Floq}$, is:
\begin{gather}
    \Lambda_{Floq} = \left( U_{A_8}^n \otimes \Bar{U}_{A_8}^n \right)  \left( U_{A_7}^n \otimes \Bar{U}_{A_7}^n \right)... \left( U_{A_1}^n \otimes \Bar{U}_{A_1}^n \right) \label{eq:lambda Floquet}
\end{gather}
where we have adapted the 5 step procedure on a square lattice of \cite{rudner2013anomalous} to an analogous 8 step procedure on a Lieb lattice. To simplify the comparison, we have neglected the 5th "holding period" step and sublattice potentials in the original Rudner et. al. procedure \cite{rudner2013anomalous}.    

Note the measurement protocol in the Zeno limit (Eq. \ref{eq:lambda final}) is precisely the Floquet evolution interspersed with measurements between each step.  Markedly, when $p=1$ the two evolutions are equivalent since the measurement projectors act trivially in the perfect switching case (when the initial G is diagonal).  




We now turn to investigate the simulated dynamics in this regime.  Away from the perfect switching cycle, $p=1$, we find an interesting distinction between the Floquet evolution and the Zeno limit of the measurement/evolution cycle as shown in Fig. \ref{fig: compare floquet and meas}. Examining the charge transfer on the resolution of the $8$ steps per cycle, we find a double step structure in the charge transfer which is not present in the corresponding Floquet evolution.  Namely, the 3rd and 8th step of the measurement protocol each contribute half of the total flow per complete cycle. The reason for this double step structure is the following.  The dynamics of particles in the lattice are governed by a classical, chiral random walk determined by $R_{cyc}$.  The 3rd and the 8th step are the only two steps that cross the slice through the Lieb lattice, and thus all transport must occur within these 2 steps.  For a particle starting far away from the slice, all information about whether the particle would cross the slice during the 3rd or 8th step in the deterministic $p=1$ case is lost.  Hence, in the long time dynamics, a particle is equally likely to cross the slice on either step leading to the observed double step structure. We emphasize that this double step structure holds for all $p \neq 1$ (see figure \ref{fig:Converging flow compare offset}).  However, similar to the Floquet evolution, the Flow per full measurement cycle decreases away from 1 for $p<1$ as shown in figure \ref{fig:analytic flow per cycle}.

\section{Away from the Zeno limit} 
\begin{figure}
    \centering
    \includegraphics[width=0.45\textwidth]{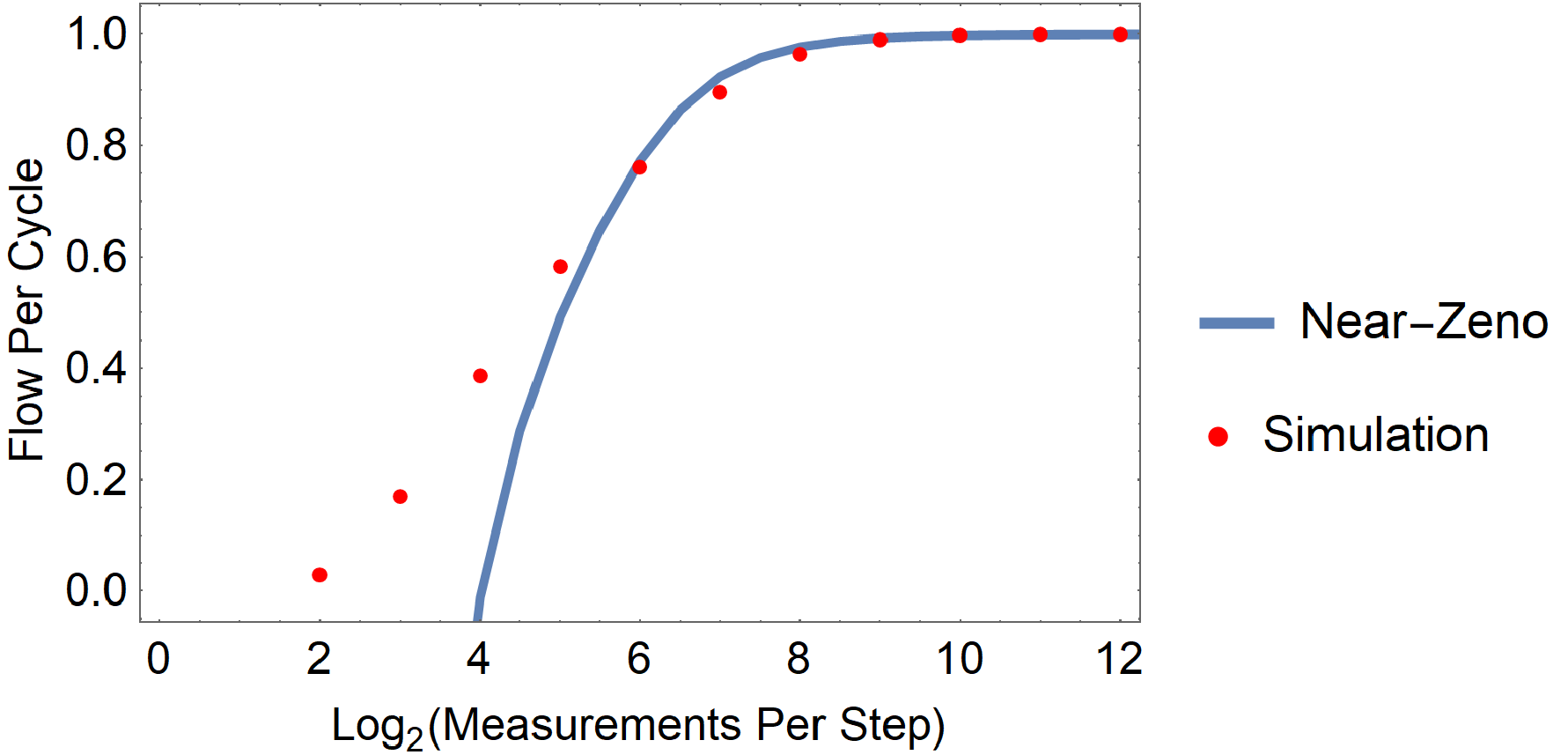}
    \caption{Flow per cycle as the measurements per step moves away from the Zeno limit.  Compared are the values found from the near-Zeno limit approximation, Eq. \eqref{eq:F final} with the transformation Eq. \eqref{eq: near-zeno R transform}, and the flow found from direct simulation.  Both analytics and simulations are done in the perfect switching cycle, i.e. $\frac{T}{8} = \frac{\pi}{2}$.}
    \label{fig:Away From Zeno Flow}
\end{figure}

We now turn to consider the important question of whether the flow is still present when the frequency of measurements is reduced, i.e. we study the evolution under our measurement protocol away from the Zeno limit.   In Fig. \ref{fig:Away From Zeno Flow} we show the flow as function of $\log(n)$. We see that the flow is reduced, but still finite as the measurement frequency is reduced, crossing over from near constant behavior at high frequency, to roughly logarithmic behavior, $F\sim 0.2\log_2(n)-0.4$ at low frequency $n$, with $F\sim 0.2$ particles per cycle at $n=8$ measurements per step.

The blue line in Fig. \ref{fig:Away From Zeno Flow}  represents an analytic perturbative near-Zeno correction which fits the simulations remarkably well for $n>64$. To arrive at it, we start with Eq. \eqref{eq:lambda final}, now retaining terms up to and including order $O(n \tau^2)$.  We prove in appendix \ref{appendix Near-Zeno} that the  resultant evolution, to order $O(n \tau^2)$,  can still be completely described in terms of the dynamics of the diagonal of $G$, with the classically stochastic transfer matrices $R_i$ replaced by the matrices $R_{nz,i}$ given by
\begin{equation}
    R_{nz,i} = R_i - n \tau^2 \Tilde{R}_i
    \label{eq: near-zeno R transform}
\end{equation}
where $\Tilde{R}_i$ is the near-Zeno correction to $R_i$.  As in the Zeno case, we define 
\begin{gather}
    R_{nz} = R_{nz,8} R_{nz,7} R_{nz,6} R_{nz,5} R_{nz,4} R_{nz,3} R_{nz,2} R_{nz,1}
    \label{eq:Rnz}
\end{gather}
and, in treating $R_{nz}$, only terms up to $O(n \tau^2)$ are kept after combining equations \eqref{eq: near-zeno R transform} and \eqref{eq:Rnz}.  
Finally, the blue line of Fig. \ref{fig:Away From Zeno Flow} is obtained by substituting \eqref{eq:Rnz} into \eqref{eq:F final}.  In Appendix \ref{appendix Near-Zeno}, we solve for \eqref{eq:Rnz} explicitly, but here we will focus only on the flow resulting from $R_{nz}$. We also note here that, similar to the Zeno limit case, the flow in the near-Zeno limit is protected to perturbations localized on the boundaries (see appendix \ref{appendix: Robustness of Flow}).  Furthermore, numerical simulations suggest that this protection persists even in the low frequency measurement regime.  We leave a detailed investigation of this observation to future work.      

\begin{figure}
    \centering
    \includegraphics[width=0.5\textwidth]{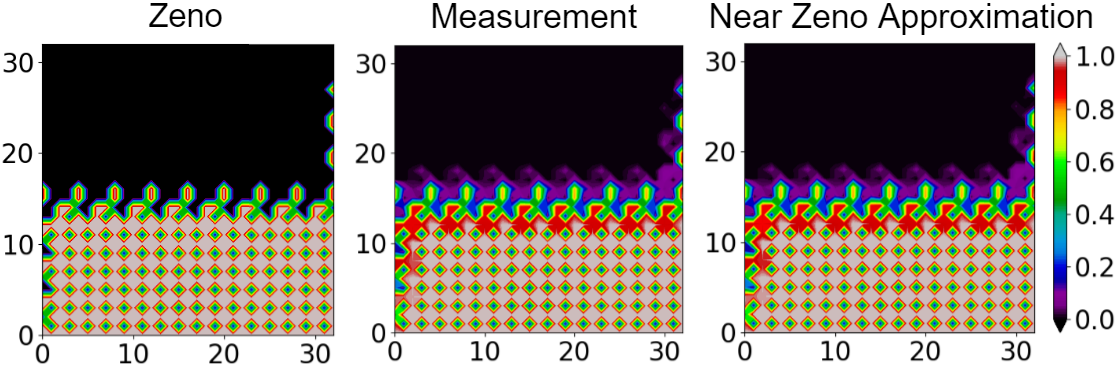}
    \caption{A comparison of, from left to right, the zeno limit, the full measurement protocol with 500 measurements per measurement step, and the near zeno approximation with 500 measurements per measurement step - all with $\frac{T  }{8} = \frac{\pi}{2}$.  Plotted are the local particle densities for a 33 by 33 site Lieb lattice after 51 measurement steps for the lower-half filled plane set up given in Fig. \ref{fig:Flow Diagram}.}
    \label{fig:near zeno compare}
\end{figure}

 In Fig \ref{fig:near zeno compare} we show what the evolution of density in the system away from the Zeno limit looks like. The main feature is clearly the ability of particles to spread faster into the bulk, since the evolution is not confined as effectively to a sequence of two-site evolution steps as in the Zeno case. We emphasize, however, that there is still significant charge transport even far away from the Zeno regime (Fig. \ref{fig:Away From Zeno Flow}).  On the other hand, the double step structure is broken with the 8th step in the measurement cycle providing an increasing percentage of the total flow per cycle as the number of measurements per measurement step is reduced.  This is shown, for example, in Fig. \ref{fig:After3 After5 Full Measure Compare}. This is because particles on the edge are less affected by the move away from the Zeno limit (as they have fewer neighboring sites to spread too).  Since the 8th measurement step hops across the Flow cut at the edge, a larger percentage of the Zeno limit flow is retained.

\section{Concluding remarks} 
In this work we presented a framework for inducing edge modes via measurement protocols. Our work is complementary to the many recent advances in studying time periodic systems such as topological Floquet insulators \cite{lindner2011floquet,rudner2013anomalous}.
The resultant behavior is a remarkable demonstration of the role of an observer in quantum mechanics as fundamentally different from a classical observer. 

Several remarks are in order regarding open problems. First, we emphasize that the behavior analyzed in this paper is that of the average transport and dynamics of densities over all possible measurement outcomes. While it is reasonable to expect that such an average would well represent the typical
behavior of the system for a typical history of measurement outcomes (a``quantum trajectory''), it is of much interest to study how well this expectation holds by studying both  fluctuations and the behavior of the quantum trajectories in our system. 

While we have concentrated on the study of the two point function $G$, it would be also interesting to establish the limiting behavior of the many body density matrix $\rho$ as the system is observed. In particular, this would allow us to study the development of entropy and non trivial correlation in the system. Indeed, in recent works, e.g. \cite{skinner2019measurement,Yaodong2018, li2019measurement, chan2019unitary}, it has been shown that certain protocols of repeated measurements interspersed with free unitary evolution induce a phase transition in the R\'{e}nyi entropy dependant on the rate of measurement.  In our model, we have found that no two-point correlations are generated up to first order in the expansion away from the Zeno limit, keeping the system close to a product state at all times.  However, for low measurement rates, these correlations are clearly generated.  This suggests phase transitions of mutual information measures with the measurement rate may be present. 


It is important to note that, while in this work we have focused mainly on the Lieb lattice, our procedure may be easily generalized to other lattices.  For example, we provide a similar 8-step protocol on a square lattice and a 6-step protocol on a ``modified'' kagome lattice in appendix \ref{appendix other lattices}.  Furthermore, we describe some restrictions on the kinds of protocols that can be implemented on a given lattice.  

We note that while our dynamics is driven by non-interacting evolution,  the formalism (see \cite{klich2019closed}) allows for an arbitrary initial state, including interesting highly correlated ones. Moreover, we expect that in the Zeno limit, the inclusion of certain interactions may be efficiently implemented with a proper modification of the current treatment, which we leave for future work.

Finally, we suggest that a measurement protocol such as ours, while challenging, may be  experimentally realizable. One possibility is the use of quantum dot arrays as the underlying lattice \cite{kagan2016building}. Another promising direction is quantum gas microscopes. Here, experiments  working with ultracold $^6$Li fermions have established  the ability to resolve particle presence at single sites see e.g. \cite{parsons2015site,Brown2019,Vijayan2020}. 


\section*{Acknowledgments}
I.K. would like to thank Kun-Woo Kim for discussions. The work of I.K., B.J.J.K. and M.W. was supported in part by the NSF grant DMR-1918207. G.R. acknowledges support  from  the  Institute  of  Quantum Information  and  Matter,  an  NSF  Physics  Frontiers  Center funded  by  the  Gordon  and  Betty  Moore  Foundation, and the Simons  Foundation, as well as to  the NSF DMR grant number 1839271.  This work was performed in part at Aspen Center for Physics, which is supported by National Science Foundation grant PHY-1607611.

. 
\bibliography{measurement_chirality.bib}
\bibliographystyle{unsrt}


\onecolumngrid
\appendix
    

\section{Closed Hierarchy Framework}\label{appendix closed hierarchy framework}
We begin with the most general evolution of a density matrix

\begin{gather}
    \rho \longrightarrow {\cal L}(\rho) = \sum_\nu A_\nu \rho A_\nu^\dagger \hbox{ ; } \sum_\nu A_\nu^\dagger A_\nu = 1
\end{gather}
This form ensures that $\rho$ remains non-negative and the normalization condition on the Krauss operators $A_\nu$ preserves $\Tr{\rho}=1$.

The evolution of a general correlation function
\begin{gather}
    \Braket{a^\dagger_{i_1}... a^\dagger_{i_{\ell_1}} a_{i_{(\ell_1 + 1)}}...a_{i_{(\ell_1 + \ell_2)}}} = \Tr{\rho a^\dagger_{i_1}... a^\dagger_{i_{\ell_1}} a_{i_{(\ell_1 + 1)}}...a_{i_{(\ell_1 + \ell_2)}}}
\end{gather}
is given by
\begin{gather}
    \Braket{a^\dagger_{i_1}... a^\dagger_{i_{\ell_1}} a_{i_{(\ell_1 + 1)}}...a_{i_{(\ell_1 + \ell_2)}}} \longrightarrow \Braket{a^\dagger_{i_1}... a^\dagger_{i_{\ell_1}} a_{i_{(\ell_1 + 1)}}...a_{i_{(\ell_1 + \ell_2)}}} + \sum_\nu \Tr{\rho A_\nu^\dagger \left[ a^\dagger_{i_1}... a^\dagger_{i_{\ell_1}} a_{i_{(\ell_1 + 1)}}...a_{i_{(\ell_1 + \ell_2)}},A_\nu \right]}
\end{gather}
where we have used the normalization condition of $A_\nu$.  Note, the $\ell_1 + \ell_2$ correlation function is taken to a, in general, higher order correlation function leading to a hierarchy of equations.  A tractable subset of this general evolution can be found by taking the two-point function $G_{i j} \equiv \Braket{a_i^\dagger a_j}$, and asking under what set of Krauss operators does the hierarchy close, i.e. $G \longrightarrow G' = {\cal K}(G)$.

In \cite{klich2019closed}, it is shown that, for fermions on a lattice, the following Krauss operators form the complete set of all possible operations that close the hierarchy on the two point function level:
\begin{subequations}
\begin{eqnarray}
     \hbox{Non-interacting Evolution: }& {\cal L}_u(\rho) = {\cal U }\rho {\cal U}^\dagger \\
     \hbox{Particle Detection: }& {\cal L}_{D,i}(\rho) = n_i \rho n_i + (1-n_i) \rho (1-n_i) \\
     \hbox{Soft Particle Injection: }& {\cal L}_{in,i,\epsilon}(\rho) = \epsilon (2-\epsilon) a_i^\dagger \rho a_i + (1-\epsilon(1-n_i)) \rho (1-\epsilon(1-n_i))\\
     \hbox{Soft Particle Extraction: }& {\cal L}_{out,i,\epsilon}(\rho) = \epsilon (2-\epsilon) a_i \rho a_i^\dagger + (1-\epsilon n_i) \rho (1-\epsilon n_i)
\end{eqnarray}
\end{subequations}
Here,  ${\cal U }$ is assumed to describe non-interacting evolution, under which, fermion operators transform as ${\cal U}^{\dag}a^{\dag}_{i} {\cal U}=U_{ ij }a^\dag_{j}$, where $U$ is called a single-particle evolution. We have also denoted 
$n_i = a_i^\dagger a_i$ the number operator, and $\epsilon$ is a real number between 0 and 1.  It is then a straight-forward task of applying the anti-commutation relations of $a^\dagger, a$ to find the corresponding transformations on the two point function:
\begin{subequations}
\begin{eqnarray}
     \hbox{Non-interacting Evolution: }& {\cal K}_U(G)_{i j} = (U G U^{\dag})_{i j} \label{eq: G free evolve}\\
     \hbox{Particle Detection: }& {\cal K}_{D,i}(G) = P_i G P_i + (1-P_i) G (1-P_i) \label{eq: G Detect}\\
     \hbox{Soft Particle Injection: }& {\cal K}_{in,i,\epsilon}(G) = (1-P_i) G (1-P_i) + (1-\epsilon) P_i G (1-P_i) + (1-\epsilon) (1-P_i) G P_i \nonumber \\
     & + (1-\epsilon)^2 P_i G P_i + \epsilon (2- \epsilon) P_i\\
     \hbox{Soft Particle Extraction: }& {\cal K}_{out,i,\epsilon}(G) = {\cal K}_{in,i,\epsilon}(G) - \epsilon (2- \epsilon) P_i 
\end{eqnarray}
\label{eq: G Transform}
\end{subequations}
Here, 
$P_i = |i \rangle \langle i |$ is the (single particle) projector onto site $i$.  

We emphasize that no approximations are used in the derivation of Eq. \eqref{eq: G Transform}.  The resulting simplicity arises completely from the restricted set of allowed Krauss operations.  Equations (\ref{eq: G free evolve},~\ref{eq: G Detect}) are the starting point for our analysis of the evolution of G in the manuscript.

\section{ Remarks about Steady States}\label{appendix remarks about steady states}
What kind of steady states can we expect in a system like ours where evolution and density measurements are intertwined? Here it is convenient to look at the steady states of the correlation matrix $G$ rather then the full density matrix $\rho$.  Let us consider how the Hilbert-Schmidt norm of $G$ changes under unitary evolution and measurements Eq. (\ref{eq: G free evolve},~\ref{eq: G Detect}) above. The Hilbert-Schmidt norm is defined as
\begin{equation}
    \norm{G}_{HS}^2 \equiv \Tr{G^\dagger G}=\sum_{ij}|G_{ij}|^2.
\end{equation}
Clearly, $\norm{G}_{HS}$, is invariant under unitary evolution of $G$.  Particle measurements of G, as described by \eqref{particle measurement transformation}, on the other hand, set to zero some of the matrix elements of $G$ and thus can only decrease $\norm{G}_{HS}$.  A necessary (though not sufficient) condition for some $G_{steady}$ to be a steady state of some super-operator $\Lambda$, i.e. $\Lambda G_{steady}=G_{steady}$ is that the Hilbert-Schmidt norm remains constant.  This provides a restriction on $\Lambda$.  Any particle detection measurement contained in $\Lambda$ must act trivially, i.e. not eliminate any matrix elements. Thus, without loss of generality writing $\Lambda \equiv \prod_i \Pi_i U_i$ we require that
\begin{equation}
    \Lambda G_{steady} = \prod_i \Pi_i U_i G_{steady} = \prod_I U_i G_{steady}
    \label{eq: Steady Requirement}
\end{equation}

Note, for our measurement procedure, this is clearly true for any scalar matrix $G_{steady}$.  For a $G_{steady}$ with a non-uniform diagonal (such as that of a single localized particle) to be a steady state of the measurement protocol, we can only satisfy Eq. \eqref{eq: Steady Requirement} in the Zeno limit with $\frac{T  }{8}$ fine tuned to $\frac{\pi}{2}$. 

One possibility to find non-equilibrium steady states in the system, as well as offer an insight into larger systems is to use particle injection and removal as was previously done in \cite{klich2019closed}. To stabilize the system where the left half is filled with particles, we may use a strip of width $L$, where we start where we constantly try to inject particles from the left, and extract any particle that arrives to the right of the sample.

In the context of the present paper, we instead look at the effective behavior of the system, when it is partially filled and evolve over times which are long, but short compared to the time it would take to arrive at the real uniform density steady state.

\section{The measurement protocol on other lattices}\label{appendix other lattices}

\begin{figure}[h]
    \centering
    \includegraphics[width=0.7\textwidth]{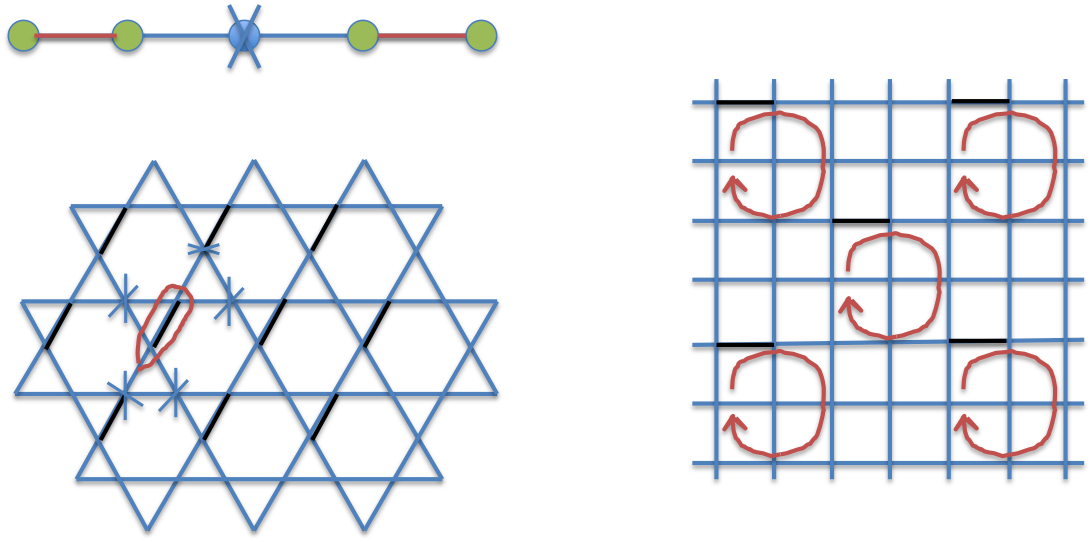}
    \caption{\textbf{Upper Left:} The measurement protocols require the bonds (red) between unmeasured sites (green) to be separated by at least two edges. This allows for at least one measured site (crossed) between them.  \textbf{Lower Left:} The naive attempt to perform the measurement protocol on the Kagome lattice does not work because two of the surrounding measured sites around the unmeasured sites (circled with red) overlap with other unmeasured sites (denoted as the ends of black links). \textbf{Right:} An example of a measurement protocol on a square lattice that satisfies the requirement that the edge distance between unmeasured pairs must be at least $2$.  If the unactivated (always measured) sites in this protocol are removed, we have exactly the 8 step protocol on a Lieb lattice introduced in this paper.}
    \label{fig:lattice_req}
\end{figure}

In this section we remark on lattices on which one can perform the measurement protocol outlined above. 
Our protocol is directly inspired by Floquet cycles where a collection of pairs of neighbouring sites are activated at any given step. To mimic this type of dynamics, we require the ability to isolate the activated pairs by performing rapid  measurements on neighbouring sites. Thus, to apply our protocol directly, we require that there is no hopping amplitude to go between two distinct pairs. For a Hamiltonian describing nearest neighbour hopping on a lattice, this means that the edge distance between unmeasured pairs is at least two (see upper left figure in Figure \ref{fig:lattice_req}).

\begin{figure}
    \centering
    \includegraphics[width=0.65\textwidth]{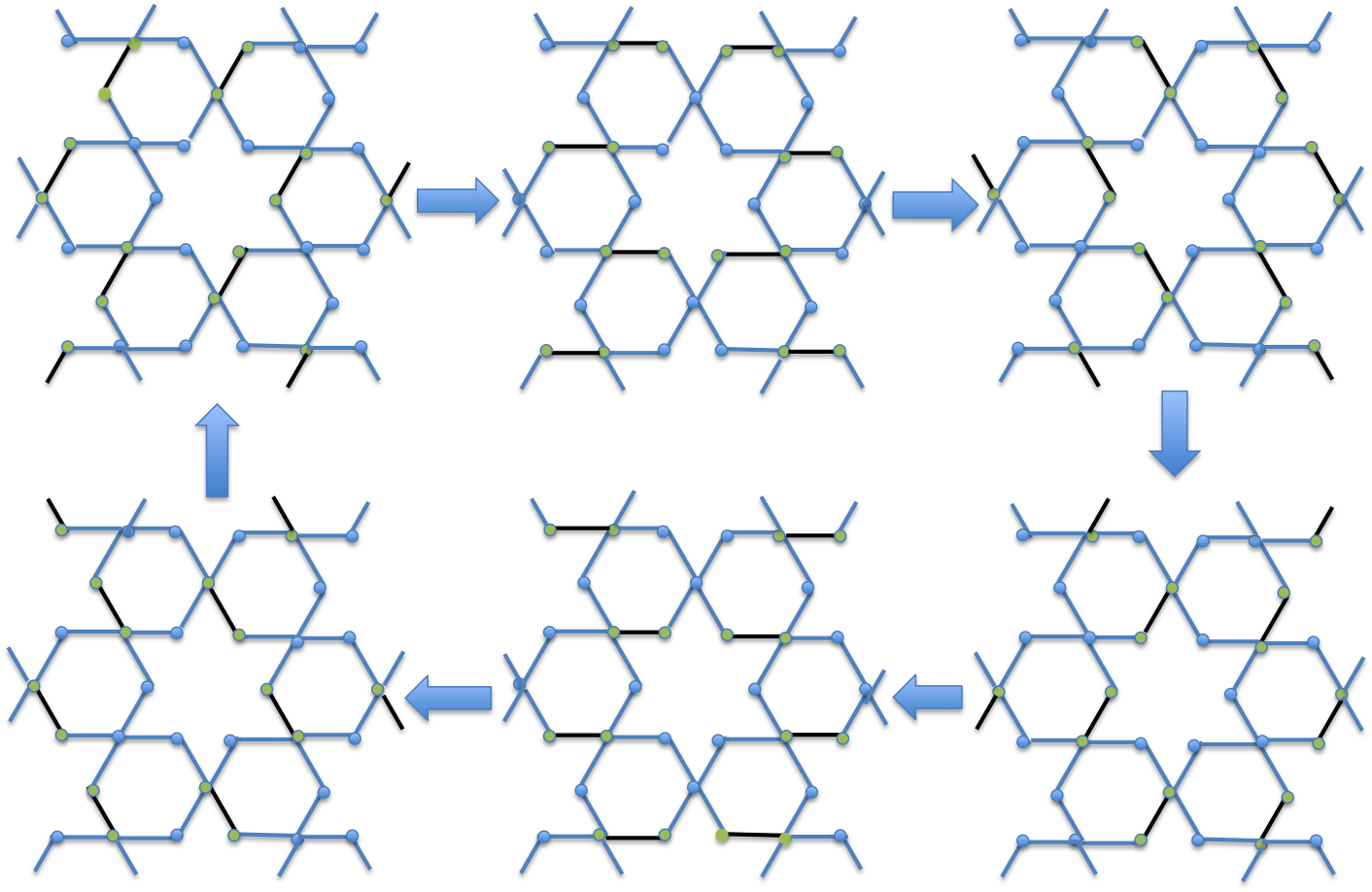}
    \caption{An example measurement protocol on a ``Modified'' Kagome lattice that utilizes $6$ measurement steps (as opposed to the $8$ step procedure used on the Lieb lattice).  The black bonds indicate free hopping pairs and measurement is indicated by blue colored sites.}
    \label{fig:modified_kogome_measurement}
\end{figure}

This restriction then rules out the simple cycle on a square lattice originally introduced in \cite{rudner2013anomalous}, where individual squares are traced out in 4 steps, as in this case the edge distance between isolated pairs is only one.  This does not, however, mean a measurement protocol cannot be implemented on a square lattice.  A solution is to increase the size of the cycle to an 8 step process that traces out a path around clusters of 4 squares (see right figure in Figure \ref{fig:lattice_req}).  Here, the edge distance between activated pairs is 3, and thus they can be isolated using rapid measurements.  Note, in this example protocol, there is a site at the center of the cycle that is never activated, i.e. always measured.  If this site is removed, we find precisely the 8 step protocol on a Lieb lattice introduced in this paper.  This choice was made to minimize the number of required measurements and to remove any spreading of particles through these unactivated sites away from the Zeno limit.  We also here give an example of another measurement protocol with $6$ measurement steps on a ``Modified'' Kagome Lattice, as opposed to the $8$ steps for our protocol on a Lieb lattice, as shown in Fig. \ref{fig:modified_kogome_measurement}

\section{Some derivation details}\label{appendix some derivation details}
In this section we supply a few more details about the formulas used in the main text.\\ 
{\bf I.} Proof of Eq. \eqref{productOfprojections}: \begin{eqnarray}
    & \prod_{a \in A^c} \pi_{a}  \nonumber= \sum_{a \in A^c} P_a \otimes P_a + P_A \otimes P_A 
\end{eqnarray}
We first note that: if $a\neq b$ then $p_a p_b=0$, $p_a (1-p_b)=p_a$. Thus in the product 
\begin{eqnarray}
    & \prod_{a \in A^c} \pi_{a}= \prod_{a \in A^c}(P_a \otimes P_a + (1-P_a) \otimes (1-P_a) )
\end{eqnarray}
the term of the form $P_a \otimes P_a $ can only appear in a product of the form $(P_a \otimes P_a)\prod_{a' \in A^c ; a'\neq a} (1-P_{a'}) \otimes (1-P_{a'})=P_a \otimes P_a$.
Next note that: $\prod_{a \in A^c} (1-P_a)=\prod_{a \in A} P_a$. Therefore $\prod_{a \in A^c} (1-P_a) \otimes (1-P_a)=P_A \otimes P_A$.
 Combining these we get Eq. \eqref{productOfprojections}. \\
{\bf II.} Derivation of equation \eqref{eq: a relation}: 
\begin{eqnarray}
     & \Pi_{A} \left[H \otimes I - I \otimes H \right] \Pi_{A} \nonumber \\
     &= \left(\sum_{a \in A^c} P_a \otimes P_a + P_A \otimes P_A \right) 
      \left[H \otimes I - I \otimes H \right]  \left(\sum_{b \in A^c} P_b \otimes P_b + P_A \otimes P_A \right) \nonumber \\
     &  = \left( P_A \otimes P_A \right) \left[H \otimes I - I \otimes H \right] \left( P_A \otimes P_A \right)   \nonumber \\
     & = H_A \otimes P_A - P_A \otimes H_A
\end{eqnarray}
where $H_A \equiv P_A H P_A$ and we used that if $a,b\in A^{c}$ then $P_A P_a=P_A P_b=0$ and that $P_a P_b=\delta_{ab}P_a$. 
\\

\section{Bulk-Edge decomposition: Proof of formula \eqref{eq:F final}.}\label{appendix bulk-edge}
Our starting point is Eq. \eqref{currentformula}.  
Taking the $\partial_\theta$ derivative and using the doubly-stochastic nature of $R_{cyc}$ (when $\theta=0$), i.e. $\bra{\mathbf{I}} R_{cyc}(\theta=0) = \bra{\mathbf{I}}$ and $R_{cyc}(\theta=0)  \ket{\mathbf{I}} = \ket{\mathbf{I}}$, we find 
\begin{eqnarray}\label{eq:FN1}
   &   F_N ={1\over N L_x}\left.i \partial _{\theta }  \chi_N(\theta) \right|_{\theta=0}={1\over N L_x } i \partial _{\theta } \langle
  \mathbf{I}|R_{cyc}^N (\theta)G\ket{ \mathbf{I}}|_{\theta=0}= \frac{1}{N L_x} \sum_{m=0}^{N-1} \langle
   \mathbf{I}|J R_{cyc}^m (\theta=0) G|\mathbf{I}\rangle \nonumber \\
  &  = \frac{1}{N L_x} \sum_{m=0}^{N-1} \langle
   \mathbf{I}|J \left[ R_{cyc}^m(\theta=0), G \right] |\mathbf{I}\rangle +\frac{1}{L_x}  \langle
   \mathbf{I}|J G |\mathbf{I}\rangle,
\end{eqnarray}
where we have defined $J= -i \partial_{\theta} R_{cyc}(\theta) |_{\theta = 0}$ and $G$ is a diagonal matrix representing the initial density distribution, i.e., if written in matrix elements,  $G_{\alpha,\beta}(\mathbf{r},\mathbf{r'}) = \delta_{\alpha \beta} \delta_{\mathbf{r},\mathbf{r'}} g_\alpha (\mathbf{r})$ with $\mathbf{r}=(x,y)$ and $\mathbf{r'}$ coordinates of the unit cell, $\alpha, \beta$ internal sites, and $g_\alpha (\mathbf{r})$ the initial probability for a particle at a site indexed by ($\mathbf{r},\alpha$). Below we suppress the angle when describing $R_{cyc}(\theta=0)$, and will just write $R_{cyc}$.

In our setup (see Fig. \ref{fig:Gm description}), we fill the system in such a way that $g_\alpha (\mathbf{r})=1$ for $y<\ell_1$ and $g_\alpha (\mathbf{r})=0$ for $y > \ell_2$. Let us define the set \begin{eqnarray}
     S_m=\{\mathbf{r}:\ell_1 - m \leq y \leq \ell_2 + m  \}.
\end{eqnarray}
The set $S_m$ contains the interface between empty and full region, "thickened" by a height $m$ below and above. Let also $P_{S_m}$ be the projection on the  set $S_m$ defined as in \eqref{defPA}. Explicitly:
\begin{equation}
 P_{S_m,\alpha,\beta}(\mathbf{r},\mathbf{r'})=\delta_{\alpha\beta} {\delta_{{\mathbf{r},\mathbf{r'}}}} 
    \begin{cases}
      1   & \ell_1 - m \leq y,y' \leq \ell_2 + m \\
        0 & \text{otherwise} \\
    \end{cases}.
\end{equation}

We now prove that we can freely move the projection $P_{S_m}$ to either side of the commutator $ \left[ R_{cyc}^m, G \right] $, namely, taking the range of $R_{cyc}$ to be short, $range(R_{cyc})\leq 1$, then
\begin{equation}\label{projectedCommutator}
    \left[ R_{cyc}^m, G \right] = P_{S_m} \left[ R_{cyc}^m, G \right]=  \left[ R_{cyc}^m, G \right] P_{S_m}=P_{S_m} \left[ R_{cyc}^m, G \right]P_{S_m}
\end{equation}

{\it Proof:}
Consider the commutator $\left[ R_{cyc}^m, G \right]$. Note that since $R_{\alpha \beta}(\mathbf{r},\mathbf{r'})=0$ if $|\mathbf{r}-\mathbf{r'}| > 1$, we have $R^m_{\alpha \beta}(\mathbf{r},\mathbf{r'})=0$ if $|\mathbf{r}-\mathbf{r'}| > m$. Therefore, looking at the matrix elements, we have $(\left[ R_{cyc}^m, G \right])_{\alpha\beta}(\mathbf{r},\mathbf{r'})=R^m_{\alpha \beta} (\mathbf{r},\mathbf{r'}) \left(g_\beta (\mathbf{r'}) - g_\alpha (\mathbf{r}) \right) = 0$ when $|\mathbf{r}-\mathbf{r'}| > m$ or $g_\beta (\mathbf{r'}) - g_\alpha (\mathbf{r})=0$. Thus, the matrix elements of $\left[ R_{cyc}^m, G \right]$ can only be non-zero when simultaneously $|\mathbf{r}-\mathbf{r'}| \leq m$ and  $g_\beta (\mathbf{r'}) - g_\alpha (\mathbf{r})\neq 0$. Let us check when the matrix elements can be non-vanishing. 

Since the system is filled the system in such a way that $g_\alpha (\mathbf{r})=1$ for $y<\ell_1$, we see that if $y<\ell_1 - m $, the  condition that $|\mathbf{r}-\mathbf{r'}| \leq m$ implies $y'\leq \ell_1$, and in particular $g_\beta (\mathbf{r'}) =g_\alpha (\mathbf{r})=1$, making the commutator vanish. Similarly, the commutator will vanish if $y>\ell_2 + m$. And of course the same considerations can be applied to $y'$. We conclude that non-zero matrix elements are only possible if 
\begin{eqnarray}
 \ell_1 - m \leq y,y' \leq \ell_2 + m    
\end{eqnarray}
which implies \eqref{projectedCommutator}. \qed

Since the boundaries of the system are not included in the $S_m$ region, we may also replace the open boundary conditions of $R_{cyc}^m$ with periodic ones, denoted by $R_B^m$, to get:
\begin{eqnarray}
      \left[ R_{cyc}^m, G \right] = P_{S_m} \left[ R_{cyc}^m, G \right] = \left[ R_{B}^m, G \right] P_{S_m}
\end{eqnarray}
Similarly, since $J$ is short ranged, far from the boundaries, the matrix elements of $J$ are identical to those of $J_B \equiv -i \partial_{\theta} R_{B}(\theta) |_{\theta = 0}$, namely $J P_{S_m}=J_B P_{S_m}$. This behavior holds when $m<min(\ell_1-range(J),L_y-\ell_2-range(J))$, which will always be assumed in the following treatment. Thus, we have:
\begin{eqnarray}
     \langle
   \mathbf{I}|J \left[ R_{cyc}^m, G \right] |\mathbf{I}\rangle= \langle
   \mathbf{I}|J P_{S_m}\left[ R_{cyc}^m, G \right] |\mathbf{I}\rangle=\langle
   \mathbf{I}|J_B P_{S_m} \left[ R_{B}^m, G \right] |\mathbf{I}\rangle 
\end{eqnarray}
Substituting in Eq. \eqref{eq:FN1} we get
\begin{gather}
    F_N = \frac{1}{N L_x} \sum_{m=0}^{N-1} \langle
   \mathbf{I}|J_B P_{S_m} \left[ R_{B}^m, G \right] |\mathbf{I}\rangle +  \frac{1}{L_x} \langle
   \mathbf{I}|J G |\mathbf{I}\rangle \nonumber \\
   = \frac{1}{N L_x} \sum_{m=0}^{N-1} \langle
   \mathbf{I}|J_B P_{S_m} R_{B}^m G |\mathbf{I}\rangle-\frac{1}{N L_x} \sum_{m=0}^{N-1} \langle
   \mathbf{I}|J_B P_{S_m} G |\mathbf{I}\rangle
   +\frac{1}{L_x} \langle
   \mathbf{I}|J G |\mathbf{I}\rangle \nonumber \\
   = \frac{1}{N L_x} \sum_{m=0}^{N-1} \langle
   \mathbf{I}|J_B P_{S_m} R_{B}^m G |\mathbf{I}\rangle+\frac{1}{N L_x} \sum_{m=0}^{N-1} \langle
   \mathbf{I}|J(G- P_{S_m} G) |\mathbf{I}\rangle
\end{gather}
where in the last line we used that in the bulk $\langle
   \mathbf{I}|J_B P_{S_m} G |\mathbf{I}\rangle=\langle
   \mathbf{I}|J P_{S_m} G |\mathbf{I}\rangle$. 
   
To proceed we note that 
\begin{eqnarray}
     G- P_{S_m} G=(1-P_{S_m})G=(P_{y< \ell_1-m}+P_{y> \ell_2+m})G=P_{y< \ell_1-m}
\end{eqnarray} 
where $P_{y < \ell_1-m}$, $P_{y> \ell_2+m}$ are projectors onto the regions with $y$ below $y=\ell_1-m$ and $y$ above $y=\ell_2+m$ respectively. Also, we used that: $P_{y < \ell_1-m}G=P_{y < \ell_1-m}$, and $P_{y < \ell_2+m}G=0$, which follow immediately from the definition of $G$. 
Therefore:
   \begin{gather} \label{eq: Bulk and almost edge}
    F_N 
   = \frac{1}{N L_x} \sum_{m=0}^{N-1} \langle
   \mathbf{I}|J_B P_{S_m} R_{B}^m G |\mathbf{I}\rangle+\frac{1}{N L_x} \sum_{m=0}^{N-1} \langle
   \mathbf{I}|J  P_{y < \ell_1-m}|\mathbf{I}\rangle
\end{gather}
 
 We can further simplify as follows. Let us assume there is no bulk current per unit cell. Then, if averaged over a bulk strip whose width is a unit cell, we have $\langle \mathbf{I}|J  (P_{y < \ell_1-m}-P_{y < \ell_1-(m-1)})|\mathbf{I}\rangle=0$, which, finally, taking $range(J)=1$, yields the form
\begin{gather}      F_N  = \frac{1}{N L_x} \sum_{m=0}^{N-1} \langle
   \mathbf{I}|J_B P_{S_m} R_{B}^m G |\mathbf{I}\rangle + \frac{1}{ L_x}\langle
   \mathbf{I}|J P_{y \leq 2} |\mathbf{I}\rangle \nonumber \\
   \equiv F_{bulk} + F_{edge}.
\end{gather}
In other words, we have split the charge transport into a term that depends only on the bulk properties of the system, 
\begin{eqnarray}
     F_{bulk}=\frac{1}{N L_x} \sum_{m=0}^{N-1} \langle
   \mathbf{I}|J_B P_{S_m} R_{B}^m G  |\mathbf{I}\rangle,\label{Fbulk}
\end{eqnarray} and a term that can be computed near the edge, \begin{eqnarray}
F_{edge} =\frac{1}{L_x} \langle
   \mathbf{I}|J P_{y \leq 2} |\mathbf{I}\rangle.
   \end{eqnarray}
   Let us consider the two terms separately.
   
{\it The edge term.}  $F_{edge}$ we can efficiently compute $\langle
   \mathbf{I}|J P_{y \leq 2} |\mathbf{I}\rangle$, which can be done explicitly by writing the transition matrix for a ladder geometry of small extension in the $y$ direction. Note that due to the short-range nature of $J$, the edge expression can be further reduced to $\langle
   \mathbf{I}|P_{y \leq 3} J P_{y \leq 2} |\mathbf{I}\rangle$. Doing so for our system on Mathematica we find with our measurement protocol $F_{edge} = p^2 + p^3 + p^4$.

   {\it The bulk term} $F_{bulk}$. Assuming the translational invariance of $R_B$, we can write $F_{bulk}$ expressed in $k$-space by defining the momentum states 
\begin{equation}
       \ket{\mathbf{k}}_\alpha = \frac{1}{\sqrt{V}} \sum_\mathbf{r} e^{i \mathbf{k} \cdot \mathbf{r}} \ket{\mathbf{r},\alpha} \hbox{ , } \ket{\mathbf{r},\alpha} = \int \frac{d^2 k}{(2 \pi)^2} e^{-i \mathbf{k} \cdot \mathbf{r}} \ket{\mathbf{k}}_\alpha
\end{equation}
where $V=L_x L_y$ is the number of unit cells. 
To proceed, let us write the uniform density vector $\ket{\mathbf{I}}$ as
\begin{equation}
       \ket{\mathbf{I}} = \sum_{\mathbf{r},\alpha} \ket{\mathbf{r},\alpha} = \sqrt{V} \sum_\alpha \ket{\mathbf{k}=0}_\alpha.
\end{equation}
Therefore, using the momentum representation in \eqref{Fbulk} we arrive at
\begin{gather}
    F_{bulk} = {V\over N L_x} \sum_{m=0}^{N-1}\sum_{\alpha \beta} (J_{B})_{\alpha \gamma} \left(\bra{\mathbf{k}=0} P_{S_m} R_B^m G \ket{\mathbf{k}=0} \right)_{\gamma \beta}
\end{gather}
To evaluate this expression, we need, explicitly
\begin{eqnarray}
    \bra{k_x=0,k_y}_\alpha G \ket{k'_x=0,k'_y}_\beta = \frac{\delta_{\alpha \beta}}{L_y} \sum_{y} { g}_\alpha (y) e^{-i y (k_y-k'_y)}.
\end{eqnarray}
where ${ g}_\alpha (y)=L_x^{-1}\sum_{x}g_{\alpha}({\bf r})$.
For the evolution, let us write $R_B^m$ in the form
\begin{eqnarray}
    \bra{k_x=0,k_y}_\alpha R_B^m \ket{k'_x=0,k'_y}_\beta 
    = \delta_{k_x k'_x}\delta_{k_y k'_y} \sum_{v=-m}^m C_{\alpha \beta m v} e^{i k_y v}.
\end{eqnarray}
where the coefficients $C_{\alpha \beta m v}$ depend on the model. The restriction $-m\leq v\leq m$ follow from the range of $R_B^m$ being limited to $m$.  
Also note that:
\begin{eqnarray}
    \bra{\mathbf{k}}_\alpha P_{S_m} \ket{\mathbf{k'}}_\beta = \frac{\delta_{\alpha \beta} \delta_{k_x k'_x}}{L_y} \sum_{\ell_1-m \leq y\leq \ell_2+m} e^{-i y (k_y-k'_y)}.
\end{eqnarray}

Putting these together we have
\begin{gather}
    \left(\bra{\mathbf{k}=0} P_{S_m}  R_B^m G \ket{\mathbf{k}=0} \right)_{\alpha \beta} \nonumber \\
    = \frac{1}{L_y} \int \frac{dk'_y}{2 \pi} \sum_{\ell_1-m \leq y\leq \ell_2+m} e^{i k'_y y} \sum_{v=-m}^m C_{\alpha \beta m v} e^{i k'_y v} \sum_{y'} g_\beta (y') e^{-i k'_y y'} \nonumber \\
    = \frac{1}{L_y} \sum_{\ell_1-m \leq y\leq \ell_2+m} \sum_{y'} \sum_{v=-m}^{m}  C_{\alpha \beta m v} \left( \int \frac{dk'_y}{2 \pi} e^{i k'_y (y-y'+v)}  \right) g_\beta (y') \nonumber \\
     =  \frac{1}{L_y} \sum_{v=-m}^{m} C_{\alpha \beta m v} \sum_{\ell_1-m \leq y\leq \ell_2+m}  g_\beta (y+v) \nonumber \\
    =  \frac{1}{L_y} \sum_{v=-m}^{m} C_{\alpha \beta m v} \sum_{\ell_1-m+v \leq y\leq \ell_2+m+v}  g_\beta (y) \nonumber \\
    = \frac{1}{L_y} \sum_{v=-m}^{m} C_{\alpha \beta m v} \left[ \left(\sum_{y=\ell_1}^{\ell_2} g_\beta (y) + m \right) - v  \right] \nonumber \\
    = \frac{1}{L_y} \left(i \partial_{k_y} [R_B^m (\mathbf{k})]_{\alpha\beta}
    +   \left(\sum_{y=\ell_1}^{\ell_2} g_\beta (y) + m \right) [R_B^m (\mathbf{k})]_{\alpha\beta} \right)|_{\mathbf{k=0}}.
    \label{eq:Fbulk two terms}
\end{gather}
Therefore, we have
\begin{gather}
    F_{bulk} = {V\over N L_x} \sum_{m=0}^{N-1}\sum_{\alpha \beta \gamma} (J_{B})_{\alpha \gamma} \left(\bra{\mathbf{k}=0} P_{S_m} R_B^m G \ket{\mathbf{k}=0} \right)_{\gamma \beta}
\nonumber \\
    =   \frac{1}{N} \sum_{m=0}^{N-1}\sum_{\alpha \beta } \left\{ i[J_B (\mathbf{k}) \partial_{k_y} (R_B^m (\mathbf{k}) )]_{\alpha\beta}|_{\mathbf{k}=0}+\left(\sum_{y=\ell_1}^{\ell_2} g_\beta (y) + m \right)[J_B (\mathbf{k}) (R_B^m (\mathbf{k}) )]_{\alpha\beta}|_{\mathbf{k}=0} \right\} \label{eq:Fbulk05} \end{gather}
Next, we note that in Eq. \eqref{eq:Fbulk05} above, we can use
\begin{eqnarray}
    \sum_{\alpha\beta} m  [J_B (\mathbf{k}) (R_B^m (\mathbf{k}) )]_{\alpha\beta}|_{\mathbf{k}=0}=0
\end{eqnarray}
This follows from the fact that $R_B$ is a stochastic matrix, with $R_{B}\ket{\mathbf{I}} = \ket{\mathbf{I}}$, and the assumption that there is no net current in the uniform density system:
\begin{eqnarray}
0=\bra{\mathbf{I}}J_B \ket{\mathbf{I}}=\bra{\mathbf{I}}J_B R_B^m \ket{\mathbf{I}}=V \sum_{\alpha\beta} [J_B (\mathbf{k}) (R_B^m (\mathbf{k}) )]_{\alpha\beta}|_{\mathbf{k}=0}=0.
     \end{eqnarray}
     Let us define $c_\alpha\equiv \sum_{y=\ell_1}^{\ell_2} g_\alpha (y)$. We then have:
  \begin{gather}
    F_{bulk} =   \frac{1}{N} \sum_{m=0}^{N-1}\sum_{\alpha \beta } \left\{ i[J_B (\mathbf{k}) \partial_{k_y} (R_B^m (\mathbf{k}) )]_{\alpha\beta}|_{\mathbf{k}=0}+[J_B (\mathbf{k}) (R_B^m (\mathbf{k}) )]_{\alpha\beta}c_{\beta}|_{\mathbf{k}=0} \right\} \label{eq:Fbulk06} \end{gather}   
     Now, using repeatedly that $R_{B}\ket{\mathbf{I}} = \ket{\mathbf{I}}$, we write :
       \begin{gather}
  \frac{1}{N} \sum_{m=0}^{N-1}  \sum_{\alpha \beta } \left\{ i[J_B (\mathbf{k}) \partial_{k_y} (R_B^m (\mathbf{k}) )]_{\alpha\beta}|_{\mathbf{k}=0} +[J_B (\mathbf{k}) (R_B^m (\mathbf{k}) )]_{\alpha\beta}c_{\beta}|_{\mathbf{k}=0} \right\} \nonumber \\ =  \frac{i}{N} \sum_{m=1}^{N-1}\sum_{\alpha \beta } \left\{  [J_B (\mathbf{k}) \sum_{q=0}^{m-1} R_B^q (\mathbf{k}) \partial_{k_y} R_B (\mathbf{k}) ]_{\alpha\beta} |_{\mathbf{k}=0}  + [J_B (\mathbf{k}) \frac{R_B(\mathbf{k})^N-I}{R_B(\mathbf{k})-I}  ]_{\alpha\beta}c_{\beta}|_{\mathbf{k}=0} \right\} \nonumber \\
    = \frac{1}{N}\sum_{\alpha \beta } \left\{ i[J_B (\mathbf{k})\Big( \frac{N [I-R_B (\mathbf{k})] + R_B^N (\mathbf{k}) - I}{[I- R_B (\mathbf{k})]^2} 
\Big) \partial_{k_y} R_B (\mathbf{k})]_{\alpha\beta}|_{\mathbf{k}=0}+[J_B (\mathbf{k}) \frac{R_B(\mathbf{k})^N-I}{R_B(\mathbf{k})-I}  ]_{\alpha\beta}c_{\beta}|_{\mathbf{k}=0} \right\}
\end{gather}
We now consider the large $N$ limit. If we assume that $c_{\beta}$ doesn't scale with $N$, the dominant term becomes
\begin{gather}
     F_{bulk} = i\sum_{\alpha\beta} [ J_B (\mathbf{k}) \frac{1}{I-R_B (\mathbf{k})}  \partial_{k_y} R_B (\mathbf{k})]_{\alpha\beta} |_{\mathbf{k}=0},
\end{gather}
which is Eq. \eqref{eq:F final bulk}.

\section{Robustness of Flow} \label{appendix: Robustness of Flow}
\begin{figure}
    \centering
    \includegraphics[width=0.75\textwidth]{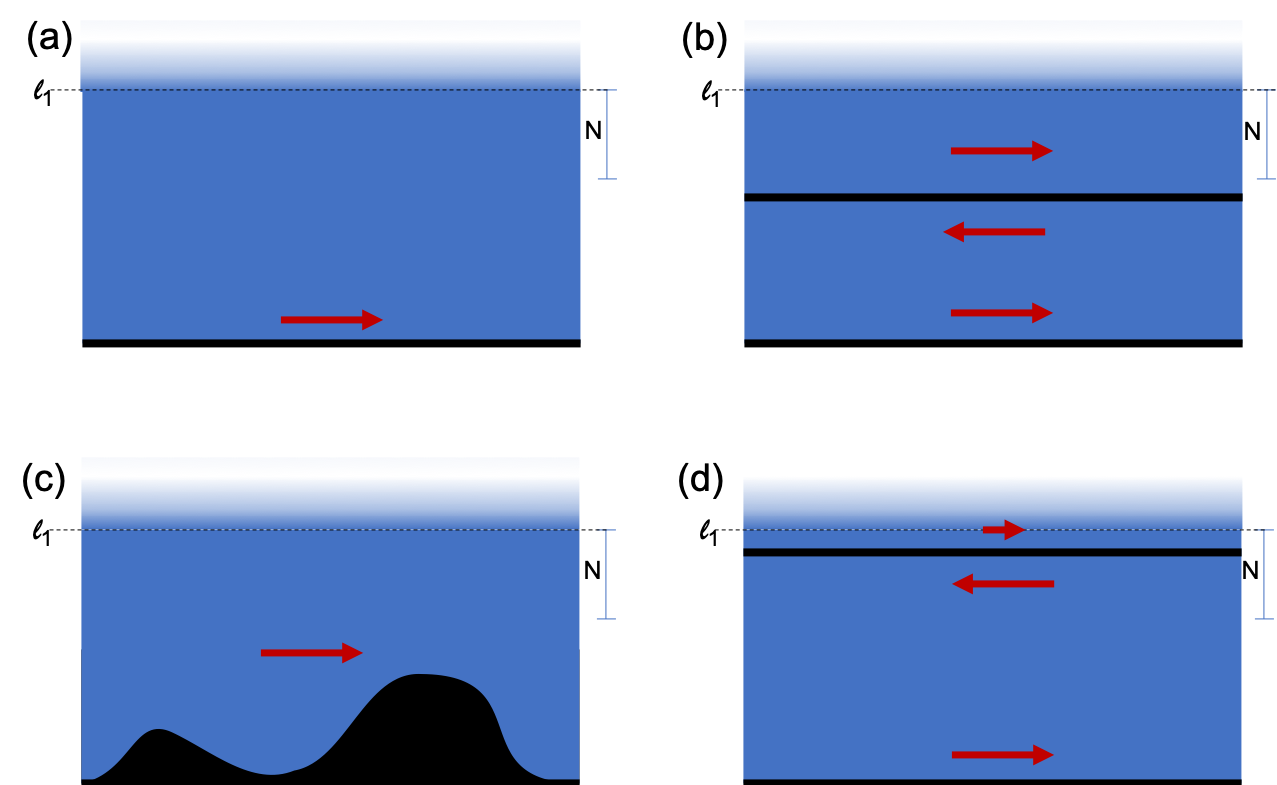}
    \caption{An illustration of the stability of the flow. When the initial state has uniform density below the line $\ell_1$, our $N$ cycle flow is only sensitive to perturbations occurring in the region above $\ell_1-N$. For large $N$, the initial density configurations (a), (b), and (c) will have the same $N$ cycle transport despite having drastic differences in R matrices (e.g. by introducing new edges in the system drawn in black above). On the other hand, panel (d) will have a reduced $N$ cycle flow, since, in contrast with panel (b), the partially filled area on the upper part of the new edge will not be enough to cancel the flow below it.}
    \label{fig:flow stability}
\end{figure}
In this section, we show that the results of Appendix \ref{appendix bulk-edge} are  robust to perturbations near the boundary (see Fig. \ref{fig:flow stability}).  Consider a perturbation of our stochastic dynamics $R_{cyc}$, affecting regions away from the bulk of the sample where we have our interface between the occupied and unoccupied regions.  Let us take it as described by a modified dynamics given by ${R}_M$ of the form:
\begin{gather}
    {R}_M (\mathbf{r},\mathbf{r'}) = 
    \begin{cases}
         \tilde{R} & r_y,r'_y \leq \ell_1 - (N+1)\\
         R_{cyc} & \ell_1 -  (N+1) \leq r_y,r'_y \leq \ell_2 + (N+1)\\
         \tilde{R}' & r_y,r'_y \geq \ell_2 +  (N+1) \\
         0 & \text{otherwise}
    \end{cases}
\end{gather}
where $\tilde{R}, \tilde{R}'$ are real matrices such that ${R}_M$ is doubly stochastic, i.e. ${R}_M$ is identical to $R_{cyc}$ in the bulk but modified near the boundary.
We now calculate the flow for this new matrix ${R}_M$ and show it is equivalent to that of $R_{cyc}$.   The situation is illustrated in figure \ref{fig:flow stability}.

Following Eq. \ref{eq:FN1}, we find the flow for ${R}_M$ is 
\begin{gather}
    F_N = \frac{1}{N L_x} i \partial_\theta \langle \mathbf{I}| {R}_M^N (\theta) G | \mathbf{I} \rangle \label{eq: FN1 general}
\end{gather}
where we have added the counting field such that $\left[{R}_M (\theta)\right]_{\alpha \beta} = \left[{R}_M \right]_{\alpha \beta} e^{-i (\beta_x - \alpha_x) \theta}$ with $\alpha = (\alpha_x,\alpha_y)$ and $\beta = (\beta_x,\beta_y)$. Defining a matrix which is only modified in the bottom edge, 
\begin{gather}
    {R_{M}'}(\mathbf{r},\mathbf{r'}) = 
    \begin{cases}
      \tilde{R} & r_y,r'_y \leq \ell_1 - (N+1)\\
      R_{cyc} & r_y,r'_y \geq \ell_1 - (N+1) \\
      0 & \text{otherwise}
    \end{cases}
\end{gather}
we note that
\begin{gather}
    {R}_{M}^N G =  R_{M}^{'N} G
    \label{eq: no top}
\end{gather}
This is because the only non-zero contributions to ${R}_M^N G$ come from terms at $r_y,r'_y \leq \ell_2 + N$, hence we are free to replace $\tilde{R}' \rightarrow R_{cyc}$ in the region $r_y,r'_y > \ell_2 + N$ without changing the result.  Combining eq. \ref{eq: FN1 general} with \ref{eq: no top} and following the rest of the steps in eq. \ref{eq:FN1} we find

\begin{gather}
    F_N = \frac{1}{N L_x} \sum_{m=0}^{N-1} \langle
   \mathbf{I}|J_M^{'} \left[ R_{M}^{'m}(\theta=0), G \right] |\mathbf{I}\rangle +\frac{1}{L_x}  \langle
   \mathbf{I}|{J^{'}_M} G |\mathbf{I}\rangle
\end{gather}
where $J^{'}_M$ is the current associated with $R_{M}^{'} $.

Similar to \ref{projectedCommutator}, we find
\begin{gather}
    \left[ R_{M}^{'m}(\theta=0), G \right] = \left[ R_{M}^{'m}(\theta=0), G \right] P_{S_m}.
\end{gather}
Note that in the region $S_m$, $R_{M}^{'}$ is identical to $R_{cyc}$.  We therefore have
\begin{gather}
    \left[ R_{M}^{'m}(\theta=0), G \right] P_{S_m} = \left[ R_{cyc}^m(\theta=0), G \right] P_{S_m} = \left[ R_{B}^m(\theta=0), G \right] P_{S_m}
\end{gather}

Repeating the steps that led to \ref{eq: Bulk and almost edge}, we find 

\begin{gather} \label{eq: general bulk and almost edge}
    F_N 
   = \frac{1}{N L_x} \sum_{m=0}^{N-1} \langle
   \mathbf{I}|J_B P_{S_m} R_{B}^m G |\mathbf{I}\rangle+\frac{1}{N L_x} \sum_{m=0}^{N-1} \langle
   \mathbf{I}|{J}_M^{'}  P_{y < \ell_1-m}|\mathbf{I}\rangle
\end{gather}
Note, the first term in \ref{eq: general bulk and almost edge} is equivalent to the $F_{bulk}$ contribution for $R_{cyc}$.  We will now show that $\frac{1}{N L_x} \sum_{m=0}^{N-1} \langle
   \mathbf{I}|{J}_M^{'}  P_{y < \ell_1-m}|\mathbf{I}\rangle$ is equivalent to the $F_{edge}$ contribution from $R_{cyc}$.  
   
Note,
\begin{gather}
    \frac{1}{N L_x} \sum_{m=0}^{N-1} \langle
   \mathbf{I}|{J}_M^{'}  P_{y < \ell_1-m}|\mathbf{I}\rangle = \frac{1}{N L_x} \sum_{m=0}^{N-1} \langle
   \mathbf{I}|{J}_M^{'}  P_{\ell_1 - N + m >y > \ell_1-N}|\mathbf{I}\rangle + \frac{1}{L_x} \langle
   \mathbf{I}|{J}_M^{'}  P_{y < \ell_1-N}|\mathbf{I}\rangle \nonumber \\
   = \frac{1}{N L_x} \sum_{m=0}^{N-1} \langle
   \mathbf{I}|J   P_{\ell_1 - N + m >y > \ell_1-N}|\mathbf{I}\rangle + \frac{1}{L_x} \langle
   \mathbf{I}|{J}_M^{'}  P_{y < \ell_1-N}|\mathbf{I}\rangle \\
   = \frac{1}{L_x} \langle
   \mathbf{I}|{J}_M^{'}  P_{y < \ell_1-N}|\mathbf{I}\rangle \nonumber 
\end{gather}
where in the second and third lines we have used the fact that ${J}_M^{'}$ is identical to $J$ for $y > \ell_1 - N$ and that $R_{cyc}$ has no bulk transport implies $\langle
   \mathbf{I}|J  P_{\ell_1 - N + m >y > \ell_1-N}|\mathbf{I}\rangle = 0$.
Furthermore,
\begin{gather}
    \frac{1}{L_x} \langle
   \mathbf{I}|{J}_M^{'}  P_{y < \ell_1-N}|\mathbf{I}\rangle = \frac{1}{L_x} \langle
   \mathbf{I}|{J}_M^{'}  \left(I - P_{y > \ell_1-N} \right)|\mathbf{I}\rangle 
\end{gather}

We now restrict ourselves to the case where $\langle
   \mathbf{I}|{J}_M^{'} |\mathbf{I}\rangle = 0$, i.e. no net current in the uniform density state.  Note, this is the case when ${R}$ is a product of bi-stochastic symmetric matrices, which includes many of the most natural perturbations near the boundary (random potentials, removed sites, variation in hopping amplitude or measurement step timing, etc.). In this case, we find
\begin{gather}
   \frac{1}{L_x} \langle
   \mathbf{I}|{J}_M^{'} \left(I - P_{y > \ell_1-N} \right)|\mathbf{I}\rangle = - \frac{1}{L_x} \langle
   \mathbf{I}|{J}_M^{'}  P_{y > \ell_1-N}|\mathbf{I}\rangle = - \frac{1}{L_x} \langle
   \mathbf{I}|J  P_{y > \ell_1-N}|\mathbf{I}\rangle \\
   = \frac{1}{L_x} \langle
   \mathbf{I}|J P_{y \leq 2} |\mathbf{I}\rangle = F_{edge} \nonumber
\end{gather}

We thus have that flow is unaffected by arbitrary evolution near the boundary.  It is only dependant on the bulk properties of the evolution.  Note, this argument also holds if $R_{cyc}$ is replaced by $R_{nz}$, the dynamics in the near Zeno case.  In other words, transport is completely protected even (to first order) away from the Zeno limit.  In fact, numerical simulations suggest that edge transport is unaffected by perturbations near the boundary even in the low frequency measurement regime.  Proof of this, however, is still a work in progress.

\section{The Near-Zeno Approximation: Derivation of $R_{nz}$} \label{appendix Near-Zeno}

 Our starting point is Eq. \eqref{eq: one meas step}.  Let us now include terms of order up to $O(\tau^2)$, and rewrite it as 
\begin{eqnarray}
     & \Pi_{A_i} (U \otimes \Bar{U}) \Pi_{A_i} = \Pi_{A_i} - i \tau \left[H_{A_i} \otimes P_{A_i} - P_{A_i} \otimes H_{A_i} \right]  - \frac{\tau^2}{2} \Pi_{A_i} \left[H \otimes I - I \otimes H \right]^2 \Pi_{A_i}+ O(\tau^3) \nonumber \\
     & = \Pi_{A_i} (U_{A_i} \otimes \Bar{U}_{A_i}) \Pi_{A_i}- \tau^2 \zeta_{A_i}(H) + O(\tau^3)
\end{eqnarray}
where 
\begin{gather}
     \zeta_{A_i}(H)=\frac{1}{2}\Pi_{A_i} \left[H^2 \otimes I + I \otimes H^2 - 2 H \otimes H \right] \Pi_{A_i}
      - \frac{1}{2} \left[H_{A_i}^2 \otimes P_{A_i} + P_{A_i} \otimes H_{A_i}^2 - 2 H_{A_i} \otimes H_{A_i} \right]
\end{gather}
From this we find
\begin{gather}
      \Pi_{A_{i+1}} \left(\Pi_{A_i} (U \otimes \Bar{U}) \Pi_{A_i} \right)^n \Pi_{A_{i-1}} \nonumber \\
      = \Pi_{A_{i+1}} \left(\Pi_{A_i} (U_{A_i} \otimes \Bar{U}_{A_i}) \Pi_{A_i}- \tau^2 \zeta_{A_i}(H) \right)^n \Pi_{A_{i-1}} + O(n \tau^3) \nonumber \\
       =\Pi_{A_i \cap A_{i+1}} (U_{A_i}^n \otimes {\Bar{U}_{A_i}}^{n}) \Pi_{A_i \cap A_{i-1}}
       - \tau^2 \Pi_{A_i \cap A_{i+1}} \sum_{m=0}^{n-1} (U_{A_i}^m \otimes \Bar{U}_{A_i}^m)  \zeta_{A_i}(H) (U_{A_i}^{n-1-m} \otimes \Bar{U}_{A_i}^{n-1-m}) \Pi_{A_i \cap A_{i-1}} + O(\tau^3 n)
      \label{eq: NearZeno}
\end{gather}

The first term in \eqref{eq: NearZeno} corresponds to the evolution in the Zeno limit and generates the operation $R_i$ on the diagonal of $G$ (as is explained in the Zeno Limit Section).  The second term, as will be shown, corresponds to the $\tilde{R}_{i}$ operations on the diagonal of $G$.

To see this, we start by noting that the operator $\Pi_{A_i \cap A_{i+1}}$ kills the correlations between every pair of sites, unless both sites are within $A_i \cap A_{i+1}$.  Hence, off-diagonal elements of $G$ are only generated if $(U_{A_i}^m \otimes \Bar{U}_{A_i}^m)  \zeta_{A_i}(H) (U_{A_i}^{n-1-m} \otimes \Bar{U}_{A_i}^{n-1-m})$ can generate correlations between the elements of $A_i \cap A_{i+1}$.  The operators $(U_{A_i} \otimes \Bar{U}_{A_i})$ can only generate correlations within the neighboring pairs inside of $A_i$.  Now, note that the neighboring pairs within $A_i$ are separated by at least 3 edges.  Therefore, to generate correlations between the neighboring pairs using a power of $H$, i.e. $H^\nu$, we must have at least $\nu \ge 3$.  $\zeta_{A_i}(H)$, on the other hand, contains $H$ with a power of at most $2$.  It follows then that neither $\zeta_{A_i}(H)$ nor $(U_{A_i} \otimes \Bar{U}_{A_i})$ can generate correlations between the adjacent pairs in $A_i$.  Hence, any correlations generated by $(U_{A_i}^m \otimes \Bar{U}_{A_i}^m)  \zeta_{A_i}(H) (U_{A_i}^{n-1-m} \otimes \Bar{U}_{A_i}^{n-1-m})$ will be subsequently killed by $\Pi_{A_i \cap A_{i+1}}$.  We thus again have that the evolution of $G$ may be described fully by the dynamics of the diagonal of $G$.  Furthermore, we may replace $\Pi_{A_i \cap A_{i+1}}$ in \eqref{eq: NearZeno} with an operator that simply kills all correlations, namely $\sum_a P_a \otimes P_a$.

\begin{figure}
    \centering
    \includegraphics[width=0.45\textwidth]{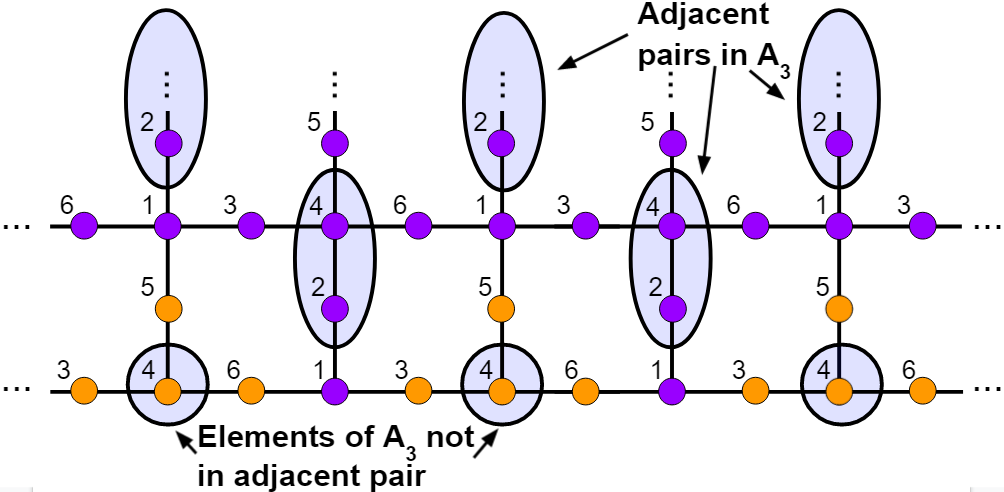}
    \caption{For any given $A_i$, here we take without loss of generality $i=3$, some elements of $A_i$ have a neighbor also in $A_i$.  Other sites in $A_i$ have no such nearest neighbor.  As described in Appendix \ref{appendix Near-Zeno}, all sites evolve in the near-Zeno approximation in one of two ways.  Lone sites in $A_i$ and nearest neighbors to lone sites in $A_i$ (as shown in orange) exhibit an evolution given by case 1 of the near-Zeno term of \eqref{eq: NearZeno}.  However, the evolution for sites that are in an adjacent pair in $A_i$ or neighboring an adjacent pair in $A_i$ (shown in purple), are governed by case 2.}
    \label{fig:Near Zeno Correction Explain}
\end{figure}

At this point in the analysis, there are two cases for the action of $(U_{A_i} \otimes \Bar{U}_{A_i})$ which we will now consider.  For sites in $A_i^c$ and for sites in $A_i$ without a neighboring site also in $A_i$ (see Fig. \ref{fig:Near Zeno Correction Explain}), $(U_{A_i} \otimes \Bar{U}_{A_i})$ simply acts as an identity.  On the other hand, for sites in $A_i$ with a neighboring site also in $A_i$, $(U_{A_i} \otimes \Bar{U}_{A_i})$ will induce Rabi oscillations within the neighboring pair inside of $A_i$.  

We further note that, for any given site $b$, the near-Zeno term in \eqref{eq: NearZeno} only induces an interaction between $b$, the closest element or pair in $A_i$ to $b$, and other nearest neighbors to this element/pair in $A_i$ (see Fig. \ref{fig:Near Zeno Correction Explain}).  This is by nature of the fact that the $H^2 \otimes I$ and $I \otimes H^2$ terms (the only terms that act non-trivially on sites outside of $A_i$) in $\zeta_{A_i}(H)$ are sandwiched by $\Pi_{A_i}$, and so can only affect nearest neighbors to any given element of $A_i$.  We therefore find that we have two disjoint sets of sites, as given in Fig. \ref{fig:Near Zeno Correction Explain}, that are affected by the near-Zeno term in \eqref{eq: NearZeno} differently.    

\textbf{Case 1:}  Here (Orange Sites in Fig. \ref{fig:Near Zeno Correction Explain}), the second term in Eq. \eqref{eq: NearZeno} becomes 

\begin{gather}
    - \tau^2 \sum_a \left( P_a \otimes P_a \right) \sum_{m=0}^{n-1}  \zeta_{A_i}(H) \sum_b \left( P_b \otimes P_b \right) \nonumber \\
    = - n \tau^2 \sum_{a,b} \left( P_a \otimes P_a \right)  \zeta_{A_i}(H) \left( P_b \otimes P_b \right)
\end{gather}

We now simplify to find

\begin{gather}
    \sum_{a,b} \left( P_a \otimes P_a \right)  \zeta_{A_i}(H) \left( P_b \otimes P_b \right) \nonumber \\
    = \sum_{a,b} \left( P_a \otimes P_a \right) \left\{ \frac{1}{2}\Pi_{A_i} \left[H^2 \otimes I + I \otimes H^2 - 2 H \otimes H \right] \Pi_{A_i}
      - \frac{1}{2} \left[H_{A_i}^2 \otimes P_{A_i} + P_{A_i} \otimes H_{A_i}^2 - 2 H_{A_i} \otimes H_{A_i} \right] \right\} \left( P_b \otimes P_b \right) \nonumber \\
      = \sum_a \deg(a) \left(P_a \otimes P_a \right) - \sum_{a,b} P_a H P_b \otimes P_a H P_b 
\end{gather}
where we have used the fact that $H_{A_i} = P_{A_i} H P_{A_i} = 0$ since in case $1$ no element of $A_i$ has a nearest neighbor also in $A_i$.

This implies that $\tilde{R}_{i}$ is given by
\begin{gather}
    \hbox{\textbf{Case 1: }} \left[ \tilde{R}_{i} \right]_{ab} = 
    \begin{cases}
      deg(a) & \text{for } a = b\\
      -1 & \text{for } a,b \text{ nearest neighbors}\\
      0 & \text{Otherwise } \\
    \end{cases}
    \label{eq: R correction 1}
\end{gather}
\textbf{Case 2:} Here (purple sites in Fig. \ref{fig:Near Zeno Correction Explain}), note that $U_{A_i} \otimes \Bar{U}_{A_i} = e^{- i \tau O}$ where we have defined 
\begin{eqnarray}
O \equiv H_{A_i} \otimes P_{A_i} - P_{A_i} \otimes H_{A_i}.      
\end{eqnarray}
Furthermore, the following relations hold
\begin{gather}
    O^2 = 2 (P_{A_i} \otimes P_{A_i} - H_{A_i} \otimes H_{A_i}) \equiv 2 E , \\
    O E = 2 O ,
\end{gather}
where we have defined $E$ in the first line and used the fact that $H_{A_i}$ simply acts like the pauli matrix $\sigma_x$ for nearest neighbors in the subspace $A_i$, i.e. $H_{A_i}^2 = P_A$.  

It then follows that

\begin{gather}
    U_{A_i} \otimes \Bar{U}_{A_i} = e^{-i \tau O} = \left(\frac{E}{2} - \frac{E}{2} \right) + 1 - 2 i \tau \frac{O}{2} - (2 \tau)^2 \frac{E}{2} + \frac{i}{3!} (2 \tau)^3 \frac{O}{2} + ... \nonumber \\
    = \left(1 - \frac{E}{2} \right) + \frac{E}{2} \cos{2 \tau} - i \frac{O}{2} \sin{2 \tau}
\end{gather}

We therefore find

\begin{gather}
    \sum_{m=0}^{n-1} (U_{A_i}^m \otimes \Bar{U}_{A_i}^m)  \zeta_{A_i}(H) (U_{A_i}^{n-1-m} \otimes \Bar{U}_{A_i}^{n-1-m}) \nonumber \\
    = \sum_{m=0}^{n-1} \left[ \left(1 - \frac{E}{2} \right) + \frac{E}{2} \cos{2 m \tau} - i \frac{O}{2} \sin{2 m \tau} \right] \zeta_{A_i}(H)  \left[ \left(1 - \frac{E}{2} \right) + \frac{E}{2} \cos{2 m \tau} + i \frac{O}{2} \sin{2 m \tau} \right] \left(U_{A_i}^{n-1} \otimes \Bar{U}_{A_i}^{n-1} \right) \nonumber \\
    = \left[ n  \left(1 - \frac{E}{2} \right) \zeta_{A_i}(H) \left(1 - \frac{E}{2} \right) + \frac{n}{2} \frac{E}{2} \zeta_{A_i}(H) \frac{E}{2} + \frac{n}{2} \frac{O}{2} \zeta_{A_i}(H) \frac{O}{2} \right] \left( 1-E \right) + O(1)
    \label{eq: near-zeno derivation}
\end{gather}

where in the last line we have restricted ourselves to the perfect switching cycle, i.e. $n \tau = \frac{\pi}{2}$, and neglected any terms in the sum that are not at least $O(n)$.

It is now convenient to rewrite $\zeta_{A_i}(H)$:

\begin{gather}
    \zeta_{A_i}(H) = \frac{1}{2} \Pi_{A_i} \left[H^2 \otimes I + I \otimes H^2 - 2 H \otimes H \right] \Pi_{A_i} - \frac{1}{2} \left[ H_{A_i}^2 \otimes P_{A_i} + P_{A_i} \otimes H_{A_i}^2 - 2 H_{A_i} \otimes H_{A_i} \right] \nonumber \\
    = \frac{1}{2} \Pi_{A_i} \left[H^2 \otimes I + I \otimes H^2 - 2 H \otimes H \right] \Pi_A - E \equiv Z - E
    \label{eq: zeta to Z - E}
\end{gather}

where $Z$ has been defined in the last line.  We may now combine Eqs. \eqref{eq: near-zeno derivation} and \eqref{eq: zeta to Z - E} to find

\begin{gather}
    \sum_{m=0}^{n-1} (U_A^m \otimes \Bar{U}_A^m)  \zeta(H) (U_A^{n-1-m} \otimes \Bar{U}_A^{n-1-m}) \nonumber \\
    = \left[ n  \left(1 - \frac{E}{2} \right) (Z-E) \left(1 - \frac{E}{2} \right) + \frac{n}{2} \frac{E}{2} (Z-E) \frac{E}{2} + \frac{n}{2} \frac{O}{2} (Z-E) \frac{O}{2} \right] \left( 1-E \right) + O(1) \nonumber \\
    = n \left[Z - \frac{E Z}{2} - \frac{Z E}{2} + \frac{3}{8} E Z E + \frac{1}{8} O Z O - E  \right] (1-E) + O(1) \nonumber \\
    = n \left[Z + E - \frac{1}{2} \{E,Z \} + \frac{1}{8} E Z E - \frac{1}{8} O Z O   \right] + O(1)
    \label{eq: simplify nz correction}
\end{gather}

where $\{E,Z \} = EZ + ZE$ is the anti-commutator.

Now, combining Eqs. \eqref{eq: NearZeno} and \eqref{eq: simplify nz correction}, we find that the near-Zeno term in \eqref{eq: NearZeno} becomes

\begin{gather}
       - n \tau^2 \sum_{a,b} \left( P_a \otimes P_a \right) \left[Z + E - \frac{1}{2} \{E,Z \} + \frac{1}{8} E Z E - \frac{1}{8} O Z O   \right] \left( P_b \otimes P_b \right)
       \label{eq: zeno + nz}
\end{gather}

Considering each of the terms in \eqref{eq: zeno + nz}, we have

\begin{eqnarray}
     \sum_{a,b} \left( P_a \otimes P_a \right) Z \left( P_b \otimes P_b \right) =& \sum_a \deg(a) \left(P_a \otimes P_a \right) - \sum_{a,b} P_a H P_b \otimes P_a H P_b \\
     \sum_{a,b} \left( P_a \otimes P_a \right) E \left( P_b \otimes P_b \right) =& \sum_{a \in A_i} P_a \otimes P_a - \sum_{a,b \in A_i} P_a H P_b \otimes P_a H P_b \\
     \sum_{a,b} \left( P_a \otimes P_a \right) \left[ - \frac{1}{2} \{E,Z \} \right] \left( P_b \otimes P_b \right) =& -\sum_{a \in A_i} \left[\deg(a) + 1 \right] \left(P_a \otimes P_a \right) + \sum_{a,b \in A_i} \left(2 + \frac{\deg(a)  + \deg(b)}{2} \right)  P_a H P_b \otimes P_a H P_b \nonumber \\
     &+ \frac{1}{2} \sum_{a \in A_i^c,b \in A_i} \left( P_a H P_b \otimes P_a H P_b + h.c. \right) \nonumber\\
     &- \frac{1}{2} \sum_{a \in A_i^c,b \in A_i} \left( P_a H H_{A_i} P_b \otimes P_a H H_{A_i} P_b + h.c. \right) \\
     \sum_{a,b} \left( P_a \otimes P_a \right) \left[ \frac{1}{8} E Z E \right] \left( P_b \otimes P_b \right) =& 2 \sum_{a \in A_i} P_a \otimes P_a - 2 \sum_{a,b \in A_i} P_a H P_b \otimes P_a H P_b \\
     \sum_{a,b} \left( P_a \otimes P_a \right) \left[ - \frac{1}{8} O Z O \right] \left( P_b \otimes P_b \right) =& - 2 \sum_{a \in A_i} P_a \otimes P_a + 2 \sum_{a,b \in A_i} P_a H P_b \otimes P_a H P_b 
\end{eqnarray}

Finally, we therefore have that $\tilde{R}_i$ becomes

\begin{gather}
    \hbox{\textbf{Case 2: }} \left[ \tilde{R}_{i} \right]_{ab} = 
    \begin{cases}
      deg(a) & \text{for } a = b \in A_i^c\\
      -1 & \text{for } a,b \in A_i^c \text{ and nearest neighbors}\\
      -\frac{1}{2} & \text{for } \left(a \in A_i \text{ and } b \text{ neighbors the adjacent pair in } A_i \text{ that includes } a \right) \text{ or vice versa}\\
      \frac{\deg(a) + \deg(b)}{2} & \text{for } a,b \in A_i \text{ and nearest neighbors}\\
      0 & \text{Otherwise } \\
    \end{cases}
    \label{eq: R correction 2}
\end{gather}

Now, Eqs. \eqref{eq: R correction 1} and \eqref{eq: R correction 2} may be combined to find the full $\tilde{R}_i$.  Note, on the seem between case 1 and case 2, for example the element $\left[ \tilde{R}_{i} \right]_{ab}$ with $a$ as an orange site in Fig. \ref{fig:Near Zeno Correction Explain} and $b$ as a blue site, case 1 and case 2 match as required for consistency.  Namely, the element $\left[ \tilde{R}_{i} \right]_{ab} = -1$ if $a,b$ are nearest neighbors, and $0$ otherwise.  Furthermore, note that $\tilde{R}_i$ is a Zero Line-Sum matrix.  Hence, the rows and columns of $R_{nz,i} = R_i - n \tau^2 \tilde{R}_i$ sum to $1$.  Furthermore, this implies the rows and columns of $R_{nz}$ also sum to $1$ as required for the usage of Eq. \eqref{eq:F final}.      

\section{Deterministic Hopping}\label{appendix deterministic hopping}
Evolution in the Zeno Limit with perfect swapping is deterministic.  Thus edge transport and bulk localization can be seen directly.

\begin{figure}[h]
    \centering
    \includegraphics[width=0.5\textwidth]{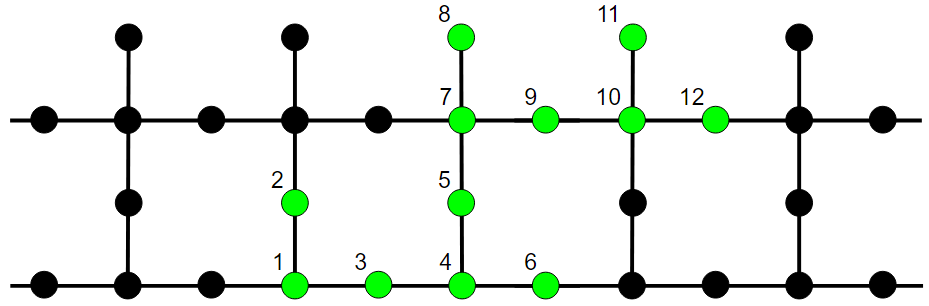}
    \caption{Lieb lattice with two layers of dynamical unit cells in the y-direction.  The bottom and top of the lattice represent a "flat" and "jagged" edge configuration respectively.}
    \label{fig:Deterministic Hopping}
\end{figure}

Figure \ref{fig:Deterministic Hopping} shows a Lieb lattice with two layers of dynamical unit cells in the y direction and infinitely many in the x direction.  The following gives the transport of a particle beginning at any given site after one complete measurement cycle (represented by arrows).  Note, after no more than 5 measurement cycles, each particle returns to either its initial position or its initial position shifted by one dynamical unit cell to the right or left. 
\\
\paragraph{Periodic Boundary Conditions:}
\begin{itemize}
    \item $1 \rightarrow 1$
    \item $2  \rightarrow 12 e^{i k_x} \rightarrow 5 \rightarrow 4 \rightarrow 3 \rightarrow 2$
    \item $6 \rightarrow 11 \rightarrow 10 \rightarrow 9 \rightarrow 8 \rightarrow 6$
    \item $7 \rightarrow 7$
\end{itemize}

where $e^{-i k_x}$ indicates a shift by one unit cell to the right.  Note that after 5 measurement cycles every particle returns to its initial position in agreement with $R_{cyc}(k,\theta)^5=I$ as described below Eq. \eqref{eq: Rcycle}.  Now turning to open boundary conditions.
\\
\paragraph{Open Boundary Conditions:}
\begin{itemize}
    \item $1 \rightarrow 6 \rightarrow 1 e^{-i k_x}$
    \item $2  \rightarrow 12 e^{i k_x} \rightarrow 5 \rightarrow 4 \rightarrow 3 \rightarrow 2$
    \item $7 \rightarrow 7$
    \item $8 \rightarrow 11 e^{i k_x} \rightarrow 10 e^{i k_x} \rightarrow 9 e^{i k_x} \rightarrow 8 e^{i k_x}$
\end{itemize}

Note here that, in contrast to the periodic boundary conditions, there is particle transport in the x direction.  Namely, particles at sites 1 and 6 shift to the right by one unit cell every 2 measurement cycles, and particles at 8, 9, 10, and 11 shift to the left one unit cell every 4 measurement cycles.

\end{document}